\documentclass[
superscriptaddress,
prc,
amsmath,
amssymb,
twocolumn,
longbibliography,
]{revtex4-1}

\usepackage{hyperref}       
\hypersetup{
    colorlinks,
    linkcolor=blue,
    citecolor=blue,
    urlcolor=blue
}
\usepackage{graphicx}
\usepackage{dcolumn}
\usepackage{bm}
\usepackage{xcolor}
\usepackage[locale=US,exponent-product=\times,per-mode=symbol,detect-all]{siunitx}

\usepackage{scalerel}
\usepackage[colorinlistoftodos]{todonotes}

\usepackage{booktabs}

\begin{document}


\title{Model comparison for initial density fluctuations in high energy heavy ion collisions}


\author{Stefan Floerchinger}
\email{stefan.floerchinger@thphys.uni-heidelberg.de}
\affiliation{Institut f\"{u}r Theoretische Physik, Universit\"{a}t Heidelberg, Philosophenweg 16, 69120 Heidelberg, Germany}
\author{Eduardo Grossi}
\email{eduardo.grossi@stonybrook.edu}
\affiliation{Department of Physics and Astronomy, Stony Brook University, Stony Brook, New York 11794, USA}%
\author{Kianusch Vahid Yousefnia}
\email{k.vahid@campus.lmu.de}
\affiliation{Institut f\"{u}r Theoretische Physik, Universit\"{a}t Heidelberg, Philosophenweg 16, 69120 Heidelberg, Germany}



\begin{abstract}
Four models for the initial conditions of a fluid dynamic description of high energy heavy ion collisions are analysed and compared. 
We study expectation values and event-by-event fluctuations in the initial transverse energy density profiles from Pb-Pb collisions. 
Specifically, introducing a Fourier-Bessel mode expansion for fluctuations, 
we determine expectation values and two-mode correlation functions of the expansion coefficients. The analytically solveable independent point-sources model is compared to an initial state model based on Glauber theory and two models based on the Color Glass Condensate framework. We find that the large wavelength modes of all investigated models show universal properties for central collisions and also discuss to which extent general properties of initial conditions can be understood analytically.
\end{abstract}


\maketitle


\section{Introduction}

Relativistic heavy-ion collisions arguably constitute one of the most spectacular physical experiments mankind is capable of conducting in a laboratory. At collider facilities like the \emph{Relativistic Heavy Ion Collider (RHIC)} and the \emph{Large Hadron Collider (LHC)}, physicists focus beams of nuclei and make them collide at relativistic energies, producing thousands of new particles per nucleus-nucleus collision. 
\cite{Abelev2013,Aamodt:2010pa, ADAMS2005102, ADCOX2005184}. 

The incident nuclei are Lorentz-contracted discs consisting of quarks and anti-quarks as well as gluons. The total energy density is maximal at the collision time of the nuclei. At this moment, these constituents of the nucleus strongly couple to each other and form a collective medium called the \emph{quark-qluon plasma (QGP)}. During LHC experiments, the energy density of this extremely dense state of matter directly after the collision is twenty times as high as that of a hadron \cite{Busza2018, Teaney:2009qa, Romatschke:2017ejr, Gale:2013da}. A droplet of QGP quickly expands and cools down. Below a critical temperature, new hadrons are formed, which is referred to as \emph{chemical freeze-out}. These particles are still interacting; only once the \emph{kinetic freeze-out} has occurred, they move freely \cite{Busza2018, Teaney:2009qa, Romatschke:2017ejr, Gale:2013da}.

In recent years, we understood that the QGP can be considered as a dissipative (and almost ideal) relativistic fluid, which offers the interesting possibility to study the relationship between microscopic properties described by \emph{quantum chromodynamics (QCD)} and macroscopic fluid fields like densities of energy or entropy. A recent numerical scheme for solving the relativistic fluid equations of motion is described in ref.\ \cite{Floerchinger:2018pje}. While these simulations can compute the time-evolution of fluid dynamic fields including their fluctuations, they need information about \emph{initial} field configurations as input. These initial conditions cannot be directly measured because experimentally accessible QGP properties like hadron spectra result from an integrated history of time-evolution, during which fluctuations in the initial conditions can be intensified or attenuated \cite{Heinz:2013th, Luzum:2013yya}. An additional complication rises from the fact that the initial conditions and in particular their symmetry properties depend on the collision centrality, i.e. whether the two nuclei collide head-on or in a peripheral manner. The underlying geometric quantity, the \emph{impact parameter}, is not directly measurable either. Most importantly, initial fields are subject to quantum fluctuations, meaning that the initial conditions vary from event to event. 

In principle, given an ensemble of initial field configurations, there are two strategies to calculate final state observables. One is to time-evolve each initial field event separately in so-called event-by-event simulations. This is numerically expensive and has the drawback that only one initial state model can be studied at a time. An alternative consists of splitting up the initial fields into a background field plus fluctuations around it and to evolve the fluctuations through \emph{response functions}. If these fluctuations are expanded in a basis of appropriately chosen modes, it suffices to solve -- just once for a given ensemble of events -- the fluid dynamic equations of motion for the background field and the response functions for the perturbations around this \cite{Floerchinger2014mode}. The response functions can actually be used to compare models of the initial state that agree in the background configuration but differ in the initial conditions for perturbations.

The present paper aims at studying the statistical properties of event-by-event initial field fluctuations in the experimentally relevant ensembles of centrality classes with randomized reaction planes. Specifically, choosing a particular mode expansion for field fluctuations, we will examine expectation values and two-mode correlation functions of the corresponding expansion coefficients. The set of basis functions constitutes Fourier- and Bessel modes and turns out to be advantageous for the decomposition of profiles that exhibit \emph{on event-average} rotation symmetry around the beam axis. 

As the initial fields of a heavy-ion collision are not directly measurable, we will have to rely on initial-condition models. A large class of them are based on \emph{Glauber theory} \cite{Miller:2007ri,Alver:2008zza}. These models consider a nucleus-nucleus collision as a superposition of independent nucleon-nucleon interactions and promote the number of nucleons participating in the collision as well as the number of binary collisions to relevant quantities. Initial event distributions are then sampled according to the position of these variables. We will make use of the initial-condition model \textsc{TrENTo}, which is based on a similar ansatz and proved to reproduce a large number of LHC experiments \cite{Moreland2015}. 

The two-mode correlation functions of Fourier-Bessel coefficients are compared to predictions of three additional initial condition models: The \emph{independent point-sources model (IPSM)} of initial conditions \cite{Alver:2008zza,Bhalerao:2006tp,Bhalerao:2011bp,Yan:2013laa,Bzdak:2013raa}. This particular model assumes fluctuations to originate from independent point-shaped contributions and allows closed-form expressions for the statistical quantities of interest. As we will see, while being limited in describing finer structures in position space, the IPSM still allows to qualitatively explain many properties of two-mode correlators.

As a further model we consider the Color Glass condensate (CGC) \cite{McLerran:1993ni, Lappi:2006hq,Lappi:2017skr, Gelis:2010nm, Albacete:2014fwa} (we concentrate on the leading approximation for large $N_c$), where two-point functions of energy density have been recently derived \cite{Albacete2019}. A variation of this model called \textsc{Magma} \cite{magma} will also be considered.

Let us mention here that all investigated models address the initial state directly after the collision. It is generally expected that a fluid description which propagates (correlations of) the energy momentum tensor to the final state becomes valid only after some period of early time non-equilibrium quantum field dynamics. During such a far-from-equilibrium phase the energy-momentum tensor and its correlation function can get modified. However, when this phase is relatively short, the modifications cannot be too large, as a consequence of relativistic causality. In terms of the mode expansion we introduce below, it is mainly the short wavelength modes corresponding to larger values of the wave numbers $m$ and $l$ that might be influenced, while one can expect that the large wavelength modes are not strongly affected, except for overall dilution as a consequence of longitudinal expansion.

This paper is structured as follows. In section~\ref{sec:mode_expansion}, we put forward our procedure of characterising fluctuations in terms of Fourier and Bessel modes. In section~\ref{sec:statistical_description} we discuss general statistical properties of the event ensembles we study. In section~\ref{sec:models} present the four initial condition models that are examined in the course of this paper. For each model, its main ideas are highlighted. Special emphasis is put on how the individual models allow to compute correlation functions. A detailed comparison of two-mode fluctuations in the four models is carried out in section~\ref{sec:comparison} and we draw conclusions in section \ref{sec:conclusions}.

\section{Mode expansion for initial field fluctuations}
\label{sec:mode_expansion}

It is convenient to express the fluid dynamic fields of interest at the initialization time in polar coordinates because they exhibit on average rotation symmetry around the beam axis (in a coordinate system that is conveniently centred).
We will express a given event profile $\epsilon(r, \phi)$ in terms of a background field plus fluctuations, and decompose the fluctuation part in terms of Fourier and Bessel modes, following the approach developed in refs. \cite{ColemanSmith:2012ka,Floerchinger:2013vua, Floerchinger2014, Floerchinger:2018pje}.

The background field will be taken to correspond to an event average. For this purpose, we introduce the following function,
\begin{equation}
W(r) = \frac{1}{2} \int_0^{2\pi} \mathrm{d}\phi\left\langle \epsilon(r, \phi) \right\rangle.
\label{eq:W}
\end{equation} 
Throughout this paper, angle brackets $\langle \cdot \rangle$ denote averages over a certain ensemble of events, for instance a centrality class. (Centrality classes will be discussed in more detail in section \ref{sec:CentralityClasses}.) We will assume that $\epsilon(r, \phi)$ is non-negative everywhere and in a given ensemble \emph{on average} normalized to unity,
\begin{equation}
\int _0^{\infty} \mathrm{d}r \, r \int_0^{2\pi} \mathrm{d}\phi \, \langle\epsilon(r, \phi)\rangle = 1.
\label{eq:ensemble_normalization}
\end{equation}
In other words, $\epsilon(r,\phi)$ corresponds to the transverse energy density divided by the corresponding integral over the transverse plane for a given ensemble or centralilty class. 

This implies that the function
\begin{equation}
\rho(r) = \sqrt{2\int_0^{r} \mathrm{d}r' \, r' W(r')}, 
\end{equation}
with
\begin{equation}
\mathrm{d}\rho\, \rho = \mathrm{d} r\, W(r) \, r,
\end{equation}
defines a map from the unbound interval $[0, \infty)$ to $[0, 1)$. The function $W(r)$ is typically non-vanishing for small radii and decays over a length scale \emph{specific} to the given centrality class, which allows a centrality-specific parametrization of the radial dependence of a given event profile in terms of $\rho$. Because it is always defined on the same interval, $\rho$ is particularly suited for the Bessel expansion to be discussed below.

The functions $\rho(r)$ and $W(r)$ define a scalar product, 
\begin{equation}
\begin{split}
(f, g) &= \int_0^{\infty} \mathrm{d}r \,r\, W(r)\, f^*(r)\, g(r) \\ 
&= \int_0^1 \mathrm{d}\rho \, \rho\, f^*(r(\rho)) \,g(r(\rho)),
\label{eq:defScalarProduct}
\end{split}
\end{equation}
which can be used conveniently to construct an orthonormal basis set. One possible choice could be a polynomial basis set constructed using the Gram-Schmidt procedure \cite{Floerchinger:2018pje}. 
Another option is to use a set of special functions that satisfy the same orthogonality properties and convenient boundary conditions, for example the cylindrical Bessel functions $J_{m}(z)$.

For $m\neq0$, let $z_l^{(m)}$ denote the $l$'th positive zero-crossing of $J_m'(z)$, the first derivative of the cylindrical Bessel function of order $m$.
Any choice of nodes will lead to the imposition of a specific boundary condition, in particular the latter corresponds to Neumann boundary conditions \cite{Floerchinger:2018pje}. 
For $m=0$ we also include the zero crossing of $J_0'(z)$ at the origin in the counting, i.e. $z_1^{(0)} = 0$.  

The set of functions $J_m(z_l^{(m)}\rho)$ are then orthogonal to each other with respect to the scalar product in eq.\ \eqref{eq:defScalarProduct}  \cite{abramowitz1965handbook},
\begin{equation}
\label{eq:bessel_orthogonality}
\int_0^{1} \mathrm{d}\rho\, \rho \, J_m\left(z_l^{(m)}\rho\right) J_m\left(z_{l'}^{(m)}\rho\right) = c_l^{(m)} \delta_{l, l'},
\end{equation}
with 
\begin{equation}
c_l^{(m)} = 
\left( (z_l^{(m)})^2 -m^2 \right)\frac{\left(J_m(z_l^{(m)}) \right)^2}{2\left( z_l^{(m)} \right)^2}, \quad\quad\quad c_1^{(0)}= \frac{1}{2}.
\end{equation}
With the case distinction in our definition of $z_l^{(m)}$, we include with $J_0(z_1^{(0)}\rho)\equiv 1$ a basis function with no positive zero-crossings and non-vanishing behavior for $\rho \to 0$. This makes our set of basis functions complete.

In summary, these considerations motivate the following mode expansion,
\begin{equation}
\begin{split}
\epsilon(r, \phi) &= \sum_{m=-\infty}^{\infty}\sum_{l=1}^{\infty} {\epsilon_{l}^{(m)} e^{im\phi} q_l^{(m)}}(r),  \\
q_l^{(m)}(r) &=
W(r)J_m(z_{l}^{(m)}\rho(r)). 
\end{split}
\label{eq:mode_expansion}
\end{equation}
Eq. \eqref{eq:mode_expansion} defines a mode expansion for the transverse energy density. Note that one can understand $m$ as an azimuthal wave number and similarly $l$ as a radial wave number with larger values corresponding to more zero crossings and therefore describing finer details in space.

The inverse relation for the coefficients $\epsilon_l^{(m)}$ is given by
\begin{equation}
\epsilon_l^{(m)} = 
\frac{1}{2\pi c_l^{(m)}}\int_0^{\infty}\mathrm{d}r  r J_m(z_l^{(m)}\rho(r))\int_0^{2\pi}\mathrm{d}\phi  e^{-im\phi} \epsilon(r, \phi).
\label{eq:back_transform}
\end{equation}
One can easily convince oneself that for real-valued fields $\epsilon(r, \phi) \in \mathbb{R}$ one has
\begin{equation}
    \epsilon_l^{(-m)} = (-1)^{m} \epsilon_l^{(m)*}.
    \label{eq:neg_coeff}
\end{equation}
Note that we have concentrated here on scalar fields, but the expansion technique can also be used in slightly modified form for vector or tensor fields such as e.\ g.\ fluid velocity and shear stress \cite{Floerchinger:2013vua}.

\section{Statistical description}
\label{sec:statistical_description}

The discussion in this section follows partly the general principles introduced in ref.\ \cite{Floerchinger2014} for a statistical characterisation of initial conditions. Let us assume that the events have been classified into centrality classes of sufficiently small extent, for example one percent. Each of these classes contains events with random orientation of the reaction plane, so that there is a \textit{statistical} azimuthal rotation symmetry. 

One may then characterise the transverse density in such a class by an expectation value 
\begin{equation}
\bar \epsilon(r) = \langle \epsilon(r,\phi) \rangle_\circ, 
\label{eq:expectationvalueSym}
\end{equation}
where we denote by $\langle \cdot \rangle_\circ$ an expectation value for an ensemble with statistical azimuthal rotation symmetry (for a more detailed discussion see section \ref{sec:Geometry} below, as well as ref.\ \cite{Floerchinger2014}).

As a consequence of the statistical rotation symmetry, the expectation value is actually independent of the azimuthal angle $\phi$. The function in eq.\ \eqref{eq:expectationvalueSym} is normalized according to eq.\ \eqref{eq:ensemble_normalization}. (This means that the overall normalization or integrated transverse energy must also be specified to characterize a given ensemble or centrality class.)  One can use $\bar \epsilon(r)$ to define the function $W(r)$ according to eq.\ \eqref{eq:W}. In terms of the Bessel-Fourier expansion in eq.\ \eqref{eq:mode_expansion} the expectation value or one-point function is characterised by the expectation value of weights 
\begin{equation}
\langle \epsilon_l^{(m)} \rangle_\circ = \frac{1}{\pi}\delta_{m, 0} \delta_{l, 1}. 
\label{eq:expValueSymmetric}
\end{equation}
In other words, for a rotation symmetric ensemble, only the $(m=0, l=1)$ coefficient has a non-vanishing expectation value, and as a consequence of the normalisation \eqref{eq:ensemble_normalization} it is given by $1/\pi$.

Fluctuations around the event-averaged profile can now be characterised in terms of correlation functions such as the connected two-point correlation function
\begin{equation}
\begin{split}
    \langle \left( \epsilon(r_1,\phi_1) - \bar \epsilon(r_1) \right)  \left( \epsilon(r_2, \phi_2) - \bar \epsilon(r_2) \right) \rangle \\ 
    = \langle  \epsilon(r_1,\phi_1)  \epsilon(r_2, \phi_2) \rangle_c.
\end{split}
    \label{eq:DefConnectedTwoPointFunction}
\end{equation}
Through eq.\ \eqref{eq:mode_expansion}, this can also be written in terms of the variance of Bessel-Fourier coefficients
\begin{equation}
    \left\langle (\epsilon^{(m_1)}_{l_1} - \langle  \epsilon^{(m_1)}_{l_1} \rangle_\circ ) ( \epsilon^{(m_2)}_{l_2} - \langle \epsilon^{(m_1)}_{l_1} \rangle_\circ ) \right\rangle = \left\langle \epsilon^{(m_1)}_{l_1}   \epsilon^{(m_2)}_{l_2} \right\rangle_c.
    \label{eq:correlationFunctionBesselFourierWeights}
\end{equation}
As a consequence of the statistical azimuthal rotation symmetry, the correlator in \eqref{eq:correlationFunctionBesselFourierWeights} is only non-vanishing when $m_1+m_2=0$.

\subsection{Geometry}
\label{sec:Geometry}
In the following sections we will investigate three initial state models for which the one-point function is used to specify the properties of the model. Specifically, for the independent point-sources model, the one-point function determines the probability distribution of sources and for the saturation models it determines the local saturation scale. For non-central collisions, the collision geometry will have a particular role. 

One can in fact describe the collision geometry in these models by introducing a \textit{non-symmetric} one-point function in a first step. It describes the expectation value of the transverse density for an ensemble with \textit{fixed} reaction plane angle $\phi_R$. For such an ensemble, the expectation value of the complex Bessel-Fourier weights is actually non-trivial and of the form
\begin{equation}
    \langle \epsilon^{(m)}_l \rangle = \bar \epsilon^{(m)}_l e^{-im \phi_R}.
    \label{eq:expValueEpsilonNonSym}
\end{equation}
The coefficients $\bar{\epsilon}_l^{(m)}$ are real-valued and non-vanishing only for even values of $m$ as a consequence of the two discrete symmetries $\phi-\phi_R \to \phi_R-\phi$ and $\phi-\phi_R \to \phi-\phi_R+\pi$. 

Note that under averaging of the reaction plane angle $\phi_R$ on the interval $[0,2\pi)$ with uniform distribution, the expression in \eqref{eq:expValueEpsilonNonSym} reduces to the one in \eqref{eq:expValueSymmetric}. In particular, all components with $m\neq 0$ are annihilated by this operation, while the $m=0$ component is unchanged. This also implies that the $m=0$ components of \eqref{eq:expValueEpsilonNonSym} are actually given by eq.\ \eqref{eq:expValueSymmetric}, i.\ e.\ $\bar \epsilon^{(0)}_l = \delta_{l,1} / \pi$.

Let us now consider two-point correlation functions. For fixed reaction plane $\phi_R$, they are of the form
\begin{equation}
    \langle \epsilon^{(m_1)}_{l_1} \epsilon^{(m_2)}_{l_2} \rangle = \langle \epsilon^{(m_1)}_{l_1} \epsilon^{(m_2)}_{l_2} \rangle_c + \bar \epsilon^{(m_1)}_{l_1} \bar \epsilon^{(m_2)}_{l_2} e^{-i (m_1+m_2) \phi_R}. 
\end{equation}
In particular, the right hand side features not only a connected part but also a disconnected one as a consequence of non-vanishing expectation values. 

If one now performs an average over the reaction plane angle $\phi_R$, one obtains for the correlation function in a rotation symmetric ensemble
\begin{equation}
\begin{split}
    \langle \epsilon^{(m_1)}_{l_1} \epsilon^{(m_2)}_{l_2} \rangle_\circ &= \frac{1}{2\pi} \int_0^{2\pi} d\phi_R \langle \epsilon^{(m_1)}_{l_1} \epsilon^{(m_2)}_{l_2} \rangle_c \\ 
    &+ \bar \epsilon^{(m_1)}_{l_1} \bar \epsilon^{(m_2)}_{l_2} \delta_{m_1+m_2,0}.
    \label{eq:TwoPointFunctSymmWithGeometryCont}
\end{split}
\end{equation}
The first part results from the averaging of the connected correlation function. 
The second term on the right hand side of \eqref{eq:TwoPointFunctSymmWithGeometryCont} arises from the geometry of the collision at non-vanishing impact parameter. It has non-trivial components in particular for $m=2$ (and more general even $m$). 

\subsection{Non-linear transformations between fields}

Some initial state models are formulated for the entropy density $s(x)$ and others for the energy density $\epsilon(x)$. In order to be able to compare them, we need to do appropriate \textit{field transformations}. In a close-to-equilibrium scenario, such a transformation can be done using the thermodynamic equations of state. A difficulty arises here because the relation between the different fields is in fact \textit{non-linear} and as such is difficult to implement in a stochastic theory. For our present purpose, it is convenient to expand the fields around some background configuration, e.\ g.\ for entropy density $s(x)=\bar s(x)+\delta s(x)$ and similarly for energy density $\varepsilon(x)= \bar \varepsilon(x)+\delta \varepsilon(x)$. Using thermal equilibrium relations in the absence of any conserved quantum numbers besides energy and momentum, one can relate the perturbations of entropy and energy density through 
\begin{equation}
    \delta \varepsilon(x) = \bar T(x) \delta s(x),
    \label{eq:relationDeltaEpsilonDeltas}
\end{equation}
where $\bar T(x)$ is the background temperature.
Using \eqref{eq:relationDeltaEpsilonDeltas} one can relate the \textit{connected} correlation functions (defined as in \eqref{eq:DefConnectedTwoPointFunction}), 

\begin{equation}
\langle \varepsilon(x)\varepsilon(y) \rangle_c = \bar T(x)\bar T(y)\, \langle s(x)s(y) \rangle_c.
\label{eq:TwoPointFunctionChangedVariables}
\end{equation}
While the fields used here are physical fields, not following the normalization condition \eqref{eq:ensemble_normalization}, it is clear that overall normalization factors can be included easily.

\section{Initial condition models}
\label{sec:models}
Having put forward a mode expansion to characterise initial fluid field fluctuations, we shall now turn to some currently popular models for the initial condition of heavy-ion collisions. For each model, we will recall how one can compute initial field configurations and characterize event averages and fluctuations in terms of correlations using the mode expansion introduced in section \ref{sec:mode_expansion}.

We will start with the \textsc{TrENTo} model, which is a Monte-Carlo implementation of a generalized Glauber model. From the numerical implementation one can obtain expectation functions and arbitrary correlation functions of transverse densities in different centrality classes.

Subsequently we will discuss the \textit{independent point-sources model} (IPSM), which is a semi-analytic model based on the assumption of strongly peaked (approximately point-like) sources that are distributed in the transverse plane according to a given probability distribution. 

Next we will illustrate the implementation of the color-glass condensate, starting from the two-point function of the energy density obtained by \cite{Albacete2019} (we concentrate on the limit of a large $N_c$) and its variation called \textsc{Magma} \cite{magma,Bhalerao:2019uzw,Giacalone:2019kgg}, in which the two-point function is simplified assuming locality in position space.

\subsection{The \textsc{TrENTo} initial condition model}

The reduced Thickness Event-by-event Nuclear Topology \textsc{(TrENTo)} initial-condition model generates event-by-event initial transverse entropy density profiles, reproducing the multiplicity distributions for a wide range of LHC experiments \cite{Moreland2015}. It constitutes a Monte Carlo model that effectively interpolates between previously existing initial condition models. \textsc{TrENTo} describes initial field profiles in terms of two nucleus thickness functions, $T_A$ and $T_B$. They are modeled as superpositions of Gaussians centered around previously sampled participating nucleon positions,
\begin{equation}
T_{A}(\vec{x}) = \sum_{i=1}^{N_\text{part}} w_{A}^{(i)} \int \mathrm{d}z \, \rho_\text{nucleon}(\vec{x} -\vec{x}_i),
\end{equation}
and similarly $T_{B}(\vec{x})$. 
The coordinates $\vec{x}_i$ denote the position of participant $i$. The strength $w_A^{(i)}$  by which a participant contributes, is sampled from a $\Gamma$-distribution with unit mean,
\begin{equation}
P_k(w) = \frac{k^k}{\Gamma(k)} w^{k-1} \exp(-kw).
\end{equation}
Here, $k>0$ is a continuous shape parameter regulating the fluctuations. The distribution has a long tail for $k<1$ while fluctuations are suppressed for $k\gg 1$.  

Given the fluctuating thickness functions of the two nuclei, the \textsc{TrENTo} model assumes the initial entropy density profile to be proportional to the generalized mean of $T_A$ and $T_B$,
\begin{equation}
\epsilon(x,y) = \mathcal{N} \left( \frac{T_{A}^p + T_{B}^p}{2}\right)^{1/p},
\label{eq:generalized_mean}
\end{equation}
with some normalization constant $\mathcal{N}$. The dimensionless parameter $p\in\mathbb{R}$ controls the mixing of the two nucleus thickness functions. Note that for $p=1$, one obtains the Glauber Monte Carlo model \cite{Miller2007}. The parameter $k$ can be tuned to match measured multiplicity distributions once $p$ has been chosen.

The code for \textsc{TrENTo} is publicly available \cite{Moreland2015}. We compute initial transverse entropy density profiles on a grid of $\num{10}\times\SI{10}{\femto\meter\squared}$ with a grid spacing of \SI{0.2}{\femto\meter} and the following parameter values,
\begin{itemize}
\item reduced thickness parameter $p=0$,
\item fluctuation parameter $k=1.4$,
\item nucleon width $\sigma=\SI{0.6}{\femto\meter}$,
\item overall normalization factor $\mathcal{N}=16$,
\item inelastic nucleon-nucleon cross section $\sigma_\text{inel}^{\text{NN}} = \SI{6.4}{\femto\meter\squared}$.
\end{itemize} 
The values for $p$, $k$, $\sigma$ and $\mathcal{N}$ have been found to best fit Pb-Pb multiplicity measurements \cite{Moreland2015}. The nucleon-nucleon cross section depends on the collision energy and has been chosen such that LHC energies of $\sqrt{s_{\text{NN}}}=\SI{2.76}{\tera\electronvolt}$ are reproduced \cite{Abelev2013}. Impact parameters are sampled from 0-\SI{20}{\femto\meter}.

With initial field profiles at hand, we can compute two-mode correlation functions of \textsc{TrENTo} profiles by numerically evaluating \eqref{eq:back_transform} and averaging over the events of a given centrality class.

While initial fields generated by the \textsc{TrENTo} model correspond to \emph{entropy} density profiles, we convert them to \emph{energy} density profiles. This allows us to compare to models that are based on energy density fields. Since the energy density scales with the power $4/3$ with the entropy density (for a thermodynamic equation of state that is approximately of the ideal gas form $p\sim T^4$), the conversion can be done by raising each \textsc{TrENTo} profile to the power of $4/3$.
Furthermore, the event distributions of a given centrality class are scaled by a common factor such that they are \emph{on average} normalized to unity, cf. \eqref{eq:ensemble_normalization}. 
In addition, each event is rotated by a random angle $\phi_R\in[0, 2\pi]$ in order to realize an ensemble of events with random orientation of the reaction plane.

\subsection{Independent point-sources model}
\label{sec:IPSM}

In contrast to \textsc{TrENTo}, the \emph{independent point-sources model (IPSM)} remarkably allows to derive analytic expressions for correlation functions \cite{Floerchinger2014}.
Let us assume that a given event profile results from $N$ independent and identically distributed contributions whose spatial extension is small compared to the system size. We will approximate them as point-like and write for the energy density
\begin{equation}
\epsilon(\vec{x}) = \frac{1}{\mu_N} \sum_{j=1}^{N} w_j \, \delta^{(2)}(\vec{x}-\vec{x}_j),
\label{eq:IPSM}
\end{equation}  
where the positions $\vec{x}_j$ are all sampled from the same probability distribution $p(\vec{x})$, normalized to unity, $\int \mathrm{d}^2x \, p(\vec{x}) = 1$. The contribution $w_j$ of each point fluctuates in strength, following a probability distribution $\tilde{p}(w)$ with unit mean and standard deviation $\sigma_w$.
In addition, we let the contribution number $N$ fluctuate according to a distribution $\hat{p}(N)$ with mean $\mu_N$ and standard deviation $\sigma_N$.

In complete analogy to ref.\ \cite{Floerchinger2014}, we can derive position space correlation functions by introducing the partition sum
\begin{align}
Z[j] &= \left\langle \exp\left(\int\mathrm{d}^2 x'\,j(\vec{x}')\epsilon(\vec{x}') \right) \right\rangle \nonumber \\
\begin{split}
&= \sum_{N}{\hat{p}(N)}\left(\prod_{j=1}^{N}\int \mathrm{d}^2x_j \,p(\vec{x_j})\int \mathrm{d}w_j\,\tilde{p}(w_j) \right) \\ &\times\exp\left(\int\mathrm{d}^2 x'\,j(\vec{x}')\epsilon(\vec{x}') \right).
\end{split}
 \label{eq:IPSMPartitionSum}
\end{align}
We obtain for the one-point function simply
\begin{equation}
\langle \epsilon(\vec{x}) \rangle = \left.\frac{\delta}{\delta j(\vec{x})}Z[j]\right|_{j=0} = p(\vec{x}).
\end{equation}
Similarly, the two-point function reads
\begin{align}
\langle \epsilon(\vec{x}) \epsilon(\vec{y}) \rangle &= 
\left.\frac{\delta^2}{\delta j(\vec{x})\delta j(\vec{y})}Z[j]\right|_{j=0}
= 
\langle \epsilon(\vec{x}) \epsilon(\vec{y}) \rangle \nonumber \\
 &= 
    ( 1- \beta)\, p(\vec{x}) p(\vec{y}) + \alpha\, p(\vec{x}) \delta^{(2)}(\vec{x}-\vec{y}).
\label{eq:two_point_space_IPSM}
\end{align}
We have introduced here the two parameters
\begin{equation}
   \alpha = \frac{1+\sigma_w^2}{\mu_N}, \quad\quad \beta = \frac{\mu_N-\sigma_N^2}{\mu_N^2}.
\end{equation}

Let the probability distribution $p(\vec{x})$ describe an event averaged field profile for fixed reaction plane angle $\phi_R$. (We will perform the averaging over $\phi_R$ later on.) We can expand then along the reaction plane angle $\phi_R$ similarly as in eq.\ \eqref{eq:mode_expansion},
\begin{equation}
p(r, \phi) = 
\sum_{\substack{m=-\infty\\ m \text{ even}}}^{\infty} \sum_{l=1}^{\infty}e^{im(\phi-\phi_R)}q_l^{(m)}(r) \bar{\epsilon}_l^{(m)}.
\label{eq:mode_expansion_p}
\end{equation}
The coefficients $\bar{\epsilon}_l^{(m)}$ are real-valued and non-vanishing only for even values of $m$ as a consequence of the two discrete symmetries $p(r, \phi_R-\phi)=p(r, \phi_R+\phi)$ and $p(r, \phi_R+\phi)=p(r, \phi_R+\phi+\pi)$.
Moreover, it follows from the event normalization (\ref{eq:ensemble_normalization}) and the definition of $W(r)$ in \eqref{eq:W} that $\bar{\epsilon}_1^{(0)}=1/\pi$ and $\bar{\epsilon}_l^{(0)}=0$ for $l>1$. 

Using the inverse relation \eqref{eq:back_transform} and the orthogonality relation \eqref{eq:bessel_orthogonality}, we can obtain two-mode correlation functions from the position space two-point function,
\begin{equation}
\begin{split}
   &  \left\langle \epsilon_{l_1}^{(m_1)}\epsilon_{l_2}^{(m_2)} \right\rangle = \frac{1}{(2\pi)^2c_{l_1}^{(m_1)}c_{l_2}^{(m_2)}}\int_{0}^{2\pi} \mathrm{d}\phi_1\, e^{-im_1\phi_1} \\ 
    & \times\int_{0}^{2\pi} \mathrm{d}\phi_2\, e^{-im_2\phi_2}  \int_{0}^{\infty} \mathrm{d}r_1 \, r_1 \int_{0}^{\infty} \mathrm{d}r_2\, r_2 \\
    & \times J_{m_1}\left(z_{l_1}^{(m_1)}\rho(r_1)\right)  J_{m_2}\left(z_{l_2}^{(m_2)}\rho(r_2)\right) \langle \epsilon(r_1,\phi_1)\epsilon(r_2, \phi_2) \rangle.
    \end{split}
    \label{eq:back_transform_2point}
\end{equation}
Performing the integrals, we find
\begin{equation}
\begin{split}
\left\langle\epsilon_{l_1}^{(m_1)}\epsilon_{l_2}^{(m_2)}\right\rangle &= \frac{(-1)^{m_1}\alpha}{2\pi^2 c_{l_1}^{(m_1)}} \delta_{l_1, l_2} \delta_{m_1+m_2,0}\\
&+ \frac{\pi^2 \alpha}{2}e^{-i(m_1+m_2)\phi_R} \\
&\times \sum_{\tilde{l}=1}^{\infty} c_{\tilde{l}}^{(m_1+m_2)} \, b_{l_1, l_2, \tilde{l}}^{(m_1, m_2, -m_1-m_2)} \bar{\epsilon}_{\tilde{l}}^{(m_1+m_2)}  \\
&+ (1 - \beta) e^{-i(m_1+m_2)\phi_R} \, \bar{\epsilon}_{l_1}^{(m_1)} \,\bar{\epsilon}_{l_2}^{(m_2)}. 
\end{split}
\label{eq:eq:two_point_IPSM}
\end{equation}
The numbers $b_{l_1, \dots, l_n}^{(m_1, \dots, m_n)}$ with $m_1+\dots+m_n = 0$ constitute integrals over Bessel functions and are defined through
\begin{equation}
\begin{split}
b_{l_1, \dots, l_n}^{(m_1, \dots, m_n)} &= \frac{1}{\pi^n} \frac{1}{c_{l_1}^{(m_1)}} \dots \frac{1}{c_{l_n}^{(m_n)}} \\ 
&\times \int_0 ^{1} \mathrm{d}\rho\,\rho\, J_{m_1}\left(z_{l_1}^{(m_1)}\rho\right) \dots J_{m_n}\left(z_{l_n}^{(m_n)}\rho\right).
\end{split}
\end{equation}

Let us now also introduce an ensemble average with a randomized reaction plane angle $\phi_R$ using the following notation,
\begin{equation}
\langle \dots \rangle_\circ = \frac{1}{2\pi} \int_0^{2\pi}\mathrm{d}\phi_R \, \langle \dots \rangle.
\end{equation}

By doing this averaging we obtain then
\begin{equation}
\begin{split}
\left\langle\epsilon_{l_1}^{(m_1)}\epsilon_{l_2}^{(m_2)}\right\rangle_\circ &= \frac{(-1)^{m_1} \alpha}{2\pi^2 c_{l_1}^{(m_1)}} \delta_{l_1, l_2} \delta_{m_1, -m_2} \\
&+ (1 - \beta) \delta_{m_1, -m_2} \, \bar{\epsilon}_{l_1}^{(m_1)} \,\bar{\epsilon}_{l_2}^{(m_2)} .
\end{split}
\label{eq:two_point_IPSM_random_general}
\end{equation}
Randomized two-mode correlators are thus only non-vanishing if $m_1+m_2=0$. This follows directly from the statistical azimuthal symmetry around the beam axis. Furthermore, two-mode correlators in the IPSM are real-valued.
From \eqref{eq:mode_expansion_p} we can read off that $\langle\epsilon_l^{(m)}\rangle = e^{-i m\phi_R}\bar{\epsilon}_l^{(m)}$. This implies
\begin{equation}
\left\langle \epsilon_l^{(m)} \right\rangle_\circ = \delta_{m, 0}\, \bar{\epsilon}_l^{(m)}.
\end{equation}
One-mode correlators of event-plane averaged events thus vanish for modes with $m\neq 0$, which is again a consequence of azimuthal rotation symmetry. Keeping in mind that $\bar{\epsilon}_1^{(0)}$ is the only non-vanishing coefficient for $m=0$ in the expansion (\ref{eq:mode_expansion_p}), we can conclude that the two-mode correlators above are equal to their respective \emph{connected} two-mode correlation function,
\begin{equation}
\begin{split}
\left \langle \epsilon_{l_1}^{(m_1)}\epsilon_{l_2}^{(m_2)} \right \rangle_\circ &= \left \langle \epsilon_{l_1}^{(m_1)}\epsilon_{l_2}^{(m_2)} \right \rangle_{\circ, c} \\ 
&= \left \langle \epsilon_{l_1}^{(m_1)}\epsilon_{l_2}^{(m_2)} \right \rangle_{\circ} - \left \langle \epsilon_{l_1}^{(m_1)} \right \rangle_{\circ}\left \langle \epsilon_{l_2}^{(m_2)} \right \rangle_{\circ},
\end{split}
\end{equation}
\emph{except} for $m_1=m_2=0$ and $l_1=l_2=1$. In the latter case, one obtains
\begin{equation}
\left \langle \epsilon_1^{(0)}\epsilon_{1}^{(0)} \right \rangle_{\circ, c}  = \frac{\alpha-\beta}{\pi^2}.
\label{eq:two_point_random_c_00}
\end{equation}

The first term in \eqref{eq:two_point_IPSM_random_general} corresponds to the connected two-mode correlation function for a spherically symmetric spatial distribution $p(r,\phi)=W(r)/\pi$, while the second term (without the part with $\beta$) accounts for geometry. Mathematically, it arises because the operations \emph{averaging over the reaction plane angle} and \emph{passing from moments to connected correlation functions} do not commute with each other \cite{Floerchinger2014}. 
Specifically the contribution due to geometry reads ,
\begin{equation}
\begin{split}
&\left\langle \epsilon_{l_1}^{(m_1)}\epsilon_{l_2}^{(m_2)} \right\rangle_{c, \text{geometry}} \\
&= 
\begin{cases}
0 & \begin{split} \text{if } (l_1, m_1, l_2, m_2)\\ = (1, 0, 1, 0), \end{split} \\
\bar{\epsilon}_{l_1}^{(m_1)}\bar{\epsilon}_{l_2}^{(m_2)}\delta_{m_1, -m_2} & \text{otherwise}.
\end{cases}
\end{split}
\label{eq:geometry}
\end{equation}

We have now fully characterized the one-point and two-point correlation functions within the independent point-sources model. By taking higher order functional derivatives of the partition sum \eqref{eq:IPSMPartitionSum} one can also calculate higher order correlation functions when needed.

\subsection{Color glass condensate large-$N_c$ model}

Recently, an analytic calculation of the connected two-point function of energy momentum tensor directly after a heavy ion collision has been reported based on the Color Glass Condensate (CGC) picture \cite{Albacete2019}. The latter is essentially a model for the field theoretic description of color fields based on the paradigm of saturation. Here we concentrate on the result of ref.\ \cite{Albacete2019} in the large $N_c$ limit and refer to the model as the \emph{CGC large-$N_c$ model}. 

Let us start with the expectation value of energy density in the McLerran-Venugopalan (MV) model \cite{Albacete2019},
\begin{equation}
\label{eq:one_point_function}
    \langle\varepsilon(\vec{x})\rangle = \frac{4}{3g^2}\bar Q_{s_1}^2(\vec{x}) \bar Q_{s_2}^2(\vec{x})\mathcal{N}^2.
\end{equation}
 The $\bar Q_{s_i}$ denote the momentum scale that characterises the colliding nuclei and $\mathcal{N}$ is a model-dependent constant. 
The latter is ultraviolet and infrared divergent for the MV model. When regularized, it reads
\begin{equation}
\mathcal{N}=\text{log}\left(
\frac{4}{m^2 L^2}
\right),
\label{eq:defNMV}
\end{equation}
where $m$ is an infrared and $1/L$ is an ultraviolet momentum regulator. Ideally one would like to consider $m\to 0$ and $L\to 0$. Because the expectation value for energy density in \eqref{eq:one_point_function} is finite, this is also the case for the product $\bar Q_s^2 \mathcal{N}$.

Note that we have been using the symbol $\varepsilon$ (instead of $\epsilon$) for the energy density, since it has not yet been normalised according to eq.\ \eqref{eq:ensemble_normalization}.
For this purpose, we introduce the scaled field $\epsilon(\vec{x})=\Lambda \varepsilon(\vec{x})$, with some normalization constant $\Lambda$ that also converts the units from $[\varepsilon]=\si{\giga\electronvolt^4}$ to $[\epsilon]=\si{\femto\meter^{-2}}$.
In order to determine $\Lambda$, consider the normalized energy density for central collisions
\begin{equation}
\left\langle\epsilon(\vec{x})\right\rangle_\circ = \frac{W(r)}{\pi},
\label{eq:center_fireball}
\end{equation}
with the background field $W(r)$ obtained from eq. (\ref{eq:one_point_function}).
With the saturation scale $Q^2_{s,0}:=\bar Q^2_s(0) \mathcal{N}$  and $W_0:=W(0)$ at the center of the fireball, we obtain
\begin{equation}
\Lambda = \frac{3 \alpha_s W_0}{ Q_{s,0}^4}; \quad \alpha_s = \frac{g^2}{4\pi}.
\label{eq:scale}
\end{equation}

Following \cite{magma, Bhalerao:2019fzp, Bhalerao:2019uzw} the saturation scale $\bar Q_{s_i}$ of a single nucleus  can be related to a thickness function, 
\begin{equation}
\label{eq:saturationscale}
 \mathcal{N} \bar Q^2_{s_i}(x) = \frac{Q^2_{s,0}}{T_A(0) } T_A(x).
\end{equation}
The thickness function is defined as \begin{equation}
    T_A(\vec x)= \int_{-\infty}^{+\infty} dz \rho_{\text{nucl}}(\vec x,z) ,
\end{equation}
with $\rho_\text{nucl}$ the nuclear charge density that can be approximated with a Woods-Saxon profile
\cite{Miller:2007ri}. Also, $Q_{s_i}(0)$ is the value of the saturation scale at the center of the nuclei.

The energy density defined in this way has to be considered at fixed impact parameter and reaction angle $\vec b$ and 
is given by 
\begin{equation}
\label{eq:CGCenergy}
\begin{split}
    \langle\epsilon(\vec{x})\rangle_{\vec b } &= \frac{4\Lambda}{3g^2}
    \frac{Q^4_{s,0}}{T_{A1}(0) T_{A2}(0)} T_{A1}\left(\vec{x}+\vec b/2\right) 
   T_{A2}\left(\vec{x}-\vec{b}/2\right)
    \end{split}
\end{equation}


The averaged energy density for each centrality class can be obtained  by averaging over a suitable distribution of impact parameters. 
For the present paper we have used the impact parameter distribution $p(b)$ obtained from \textsc{TrENTo}, since there is not a canonical choice for $p(b)$ in the CGC large-$N_c$ model itself.



The expression for the two-point function in leading order in the large $N_c$ limit as given in ref.\ \cite{Albacete2019} is 
\begin{widetext}
\begin{align}
\left\langle\epsilon(\vec{x})\epsilon(\vec{y})  \right\rangle_c =
\Lambda^2\left[\text{Cov}[\epsilon_{\scaleto{\text{\tiny MV}}{0.12cm}}](0^+; \vec x, \vec y)\right]_{N_c^0}=\Lambda^2\left[\frac{1}{g^4\,r^8}e^{-\frac{r^2}{2} \left(Q_{s\scaleto{1}{4pt}}^2+Q_{s\scaleto{2}{4pt}}^2\right)}\!\left(16+32 e^{\frac{Q_{s\scaleto{1}{4pt}}^2 r^2}{2}}\right.\right.\nonumber\\
-64e^{\frac{Q_{s\scaleto{1}{4pt}}^2 r^2}{4}}\!-4 e^{\frac{r^2}{4} \left(2 Q_{s\scaleto{1}{4pt}}^2+Q_{s\scaleto{2}{4pt}}^2\right)}\!\left( Q_{s\scaleto{2}{4pt}}^4 r^4\!-2\mathcal{N}^2\!\bar{Q}_{s\scaleto{1}{4pt}}^4 r^4+ 8\,Q_{s\scaleto{2}{4pt}}^2 r^2+48\right)\nonumber\\
+\frac{1}{8}e^{\frac{r^2}{4} \left(Q_{s\scaleto{1}{4pt}}^2+Q_{s\scaleto{2}{4pt}}^2\right)}\!\Big(Q_{s\scaleto{1}{4pt}}^4 Q_{s\scaleto{2}{4pt}}^4 r^8\!+(4\,Q_{s\scaleto{1}{4pt}}^2 Q_{s\scaleto{2}{4pt}}^2 r^6+128\,r^2)\!\left(Q_{s\scaleto{1}{4pt}}^2+Q_{s\scaleto{2}{4pt}}^2\right)\!+16\,r^4\!\left( Q_{s\scaleto{1}{4pt}}^2+ Q_{s\scaleto{2}{4pt}}^2\right)^2\!+1024\Big)\nonumber\\
\left.\left.+\,2 e^{\frac{r^2}{2} \left(Q_{s\scaleto{1}{4pt}}^2+Q_{s\scaleto{2}{4pt}}^2\right)}\!\left( \bar{Q}_{s\scaleto{1}{4pt}}^4r^4\!\left(Q_{s\scaleto{2}{4pt}}^2 r^2-4\right)\!\mathcal{N}^2\!+40\right)\right)\right]\!+\left[1\leftrightarrow2\right]\!.\label{eq:LargeN-NLO}
\end{align}
\end{widetext}
Here, $\vec x$ and $\vec y$ are positions in the transverse plane, $\vec r = \vec x - \vec y$ denotes their difference. 
The two-point function depends on the local nucleon saturation scale $Q_{s}$.
The latter is related to the thickness function  through $\bar Q_s$, the strong coupling constant $g$, and the model-dependent function $\Gamma(x)$  as in \cite{Albacete2019},
\begin{equation}
\frac{r^2 Q_s^2(\vec x,\vec y)}{4}= g^2 \frac{N_c}{2} \Gamma(\vec x-\vec y) \bar Q_s^2\left(\frac{\vec x+\vec y}{2}\right),
\end{equation}
with the function $\Gamma$ defined in the MV model as
\begin{equation}
\Gamma(r)=\frac{1}{2\pi m^2 } -\frac{r}{2\pi m}K_1(mr).
\end{equation}
As pointed out before, $\bar Q_s$ is divergent and only in combination with $\mathcal{N}$ does it lead to a finite result. Therefore it is useful to express the saturation scale directly in terms of finite quantities like the the thickness function, 
\begin{equation}
\label{eq:r2Qs2}
    \frac{r^2 Q_s^2(\vec x,\vec y)}{4}= g^2 \frac{N_c}{2} \frac{\Gamma(\vec x-\vec y) }{\mathcal{N}} T_A\left(\frac{\vec x+\vec y}{2} \right)\frac{Q^2_{s,0}}{T_A(0)}.
\end{equation}



The ratio  $\Gamma(\vec x-\vec y)/\mathcal{N}$ needs a special consideration, since for the MV model as it stands it is ill defined due to the logarithmically divergent constant $\mathcal{N}$ and the strong dependence of the function $\Gamma$ on the infrared regulator $m$. For small values of $m$  the function $\Gamma$ has the leading behaviour 
\begin{equation}
     \Gamma(r) \approx \frac{r^2}{8\pi}\text{log} \left(\frac{4}{m^2 r^2 } \right),
 \end{equation}
and together with \eqref{eq:defNMV} we find for the ratio that enters \eqref{eq:r2Qs2}, 
\begin{equation}
     \frac{\Gamma(r)}{\mathcal{N}} \approx \frac{r^2}{8\pi}\frac{\text{log} \left(\frac{4}{m^2 r^2 } \right)}{\text{log}\left(\frac{4}{m^2 L^2 } \right)}.
 \end{equation}
This depends on both infrared momentum regulator $m$ and ultraviolet momentum regulator $1/L$, respectively, while we are ultimately interested in the limit $L\to 0$ and $m\to 0$. Here we observe that if we first take $m\to 0$ we obtain
\begin{equation}
     \frac{\Gamma(r)}{\mathcal{N}} \approx \frac{r^2}{8\pi}\frac{\text{log} \left(\frac{4}{m^2 r^2 } \right)}{\text{log}\left(\frac{4}{m^2 r^2 } \right) +\text{log}\left(\frac{r^2}{L^2 } \right)}\to\frac{r^2}{8\pi},
 \end{equation}
which is finite and also independent of $L$ so that the limit $L\to 0$ can be safely taken, as well. To make progress, we assume that this is the right prescription although in principle the limits $m\to 0$ and $L\to 0$ do not need to commute. 

Possible  corrections can arise if one extends the MV model such that more scales get involved. However this goes  beyond our present scope and we therefore work with the following expression  
\begin{equation}
\label{eq}
    \frac{r^2 Q_s^2(\vec x,\vec y)}{4}= g^2 \frac{N_c}{2} \frac{r^2}{8\pi} T_A\left(\frac{\vec x+\vec y}{2} \right)\frac{Q^2_{s,0}}{T_A(0)},
\end{equation}
as the relation between the saturation scale and the nucleus thickness function. Note that $r^2$ cancels now on both sides so that the saturation scale $Q_s^2(\vec x,\vec y)$ actually depends only on $(\vec x + \vec y)/2$.



The two-point function in this model, like the one-point function, has to be considered at fixed impact parameter and reaction angle $\vec{b}$. The two-point function for a given centrality can be obtained through the impact parameter distribution $p(b)$ as a weighted average over an impact parameter window $[b_1,b_2]$.
In particular the impact parameter dependence can be introduced in a similar way as for the one-point function, namely by shifting the transverse plane dependence of the saturation scale by a $\vec{b}/2$ term. 
This corresponds to the following replacement in eq.~\eqref{eq:LargeN-NLO} 
\begin{equation}
\begin{split}
  Q^2_{s1}\to  Q^2_{s1}\left(\frac{\vec{x}+\vec{y}}{2}+\frac{\vec b}{2},\vec{x}-\vec{y}\right), \\
  Q^2_{s2}\to  Q^2_{s2}\left(\frac{\vec{x}+\vec{y}}{2}-\frac{\vec b}{2},\vec{x}-\vec{y}\right).
\end{split}
\end{equation}

The dependence of this expression on the reaction plane angle is given by the direction of $\vec{b}$, so $\phi_R$ is fixed in eq.~\eqref{eq:LargeN-NLO}. As we discussed in section \ref{sec:Geometry}, connected two-mode correlation functions in an ensemble with \emph{randomized} reaction plane angle have two contributions, one corresponding to an azimuthal average of the correlation function, and another from the product of one-point functions.




Two-mode correlation functions can be obtained using \eqref{eq:back_transform_2point} with \eqref{eq:LargeN-NLO} as position space two-point function and two additional integrations to account for the average over the impact parameter distribution $p_c(b)$ of a given ensemble and the reaction plane angle. It is very convenient to make use of some properties of the CGC large-$N_c$ position space two-point function \eqref{eq:LargeN-NLO} to reduce the number of integrals and hence the computation time. Indeed, the two-point function depends on the position variables only through
\begin{align}
r^2 &= \left|\vec{x}-\vec{y}  \right|^2 = r_1^2+r_2^2-2r_1r_2\cos(\phi_1-\phi_2),
\end{align}
as well as
\begin{align}
\begin{split}
R_\pm^2 = &   \left|\vec{x}+\vec{y}\pm \vec{b}\right|^2 = \left|\vec{x}+\vec{y}\right|^2+b^2 \\ &\pm 2b \left(r_1 \cos(\phi_1-\phi_R) + r_2 \cos(\phi_2-\phi_R) \right).
\end{split}
\end{align}
Here we parametrized the impact parameter as $\vec{b}=b (\cos(\phi_R),\sin(\phi_R))$ to make its \emph{two} independent parameters explicit. The two-point function thus takes the form
\begin{equation}
\begin{split}
    \langle \epsilon(r_1,\phi_1)\epsilon(r_2, \phi_2) \rangle_c =&\, G(r_1, r_2, b, \cos(\phi_1-\phi_2),\\ &\cos(\phi_1-\phi_R), \cos(\phi_2-\phi_R)).
\end{split}
\label{eq:Large_Nc_form}
\end{equation}

Hence, introducing additional integrations over the impact parameter and the reaction plance angle as well as performing the following change of integration variables,
\begin{equation}
 \phi_2 \rightarrow \tilde{\phi} = \phi_1-\phi_2, \quad\quad\quad \phi_R \rightarrow  \phi_2-\phi_R,
\end{equation}
expression \eqref{eq:back_transform_2point} simplifies to
\begin{align}
    \begin{split}
    \left\langle \epsilon_{l_1}^{(m_1)}\epsilon_{l_2}^{(m_2)} \right\rangle_c &=  \frac{\delta_{m_1,-m_2}}{c_{l_1}^{(m_1)}c_{l_2}^{(m_2)}} \int_{0}^{\infty} \mathrm{d}r_1 \, r_1 \int_{0}^{\infty}  \mathrm{d}r_2 \, r_2 \\
    &\times J_{m_1}\left(z_{l_1}^{(m_1)}\rho(r_1)\right) J_{m_2}\left(z_{l_2}^{(m_2)}\rho(r_2)\right)\\
    &\times \int_{0}^{\infty}\mathrm{d}b \, p_c(b) \int_{0}^{2\pi}\frac{\mathrm{d}\phi_R}{2\pi} \int_0^{2\pi}\frac{\mathrm{d}\tilde{\phi}}{2\pi}\,e^{-im_1\tilde{\phi}}  \\
    &\times
    G(r_1,r_2,b, \cos\tilde{\phi}, \cos(\tilde{\phi}+\phi_R),\cos{\phi_R}).
    \label{eq:two_mode_LargeNc}
\end{split} 
\end{align}
The Kronecker-delta arises from the $\phi_1$-integration.
Just as for the IPSM, two-mode correlators thus vanish except for $m_1+m_2= 0$. Furthermore, $G$ is upon integration over $\phi_R$ an even function in $\tilde{\phi}$ (the relative minus sign between $\tilde{\phi}$ and $\phi_R$ popping up in the fifth argument of $G$ can be handled with a change of integration variables $\phi_R\rightarrow -\phi_R$), which implies that two-mode correlators are real-valued.

We computed two-mode correlation functions for the CGC large-$N_c$ model by numerically evaluating \eqref{eq:two_mode_LargeNc} with a non-symmetric background field from (\ref{eq:one_point_function}) and adding the contribution from geometry \eqref{eq:geometry}.

\subsection{\textsc{Magma} model}
Recently, a simplification of the CGC large-$N_c$ model has been proposed. Dubbed \textsc{Magma}, it considers the energy density as a superposition of localised interactions falling off like $1/r^2$ \cite{magma}. The \emph{connected} position space two-point correlation function of energy density in this model can be written as
\begin{align}
    \langle \epsilon(\vec{x})\epsilon(\vec{y}) \rangle_c= f\left(\frac{\vec{x}+\vec{y}}{2}\right)\delta^{(2)}(\vec{x}-\vec{y}),
    \label{eq:general_form}
\end{align}
with the local function
\begin{equation}
\begin{split}
f(\vec{x}) = \frac{16\pi \Lambda^2}{9g^4}& \bar Q_{s1}^2(\vec{x})\bar Q_{s2}^2(\vec{x})\left[\bar Q_{s1}^2(\vec{x})\log\left(\frac{\bar Q_{s2}^2(\vec{x})}{m^2}\right)\right.
\\
&\left. 
+\bar Q_{s2}^2(\vec{x})\log\left(\frac{\bar Q_{s1}^2(\vec{x})}{m^2}\right)
\right].
\end{split}
\label{eq:CGC_simple}
\end{equation}
Note that we  $\Lambda^2$ to correctly scale our fields. In \eqref{eq:CGC_simple}  an infrared cutoff regulator $m$ has been introduced. It should be chosen such that the following scale hierarchy holds
\begin{equation}
    \frac{1}{\bar Q_s} \ll \frac{1}{m} \ll R,
\end{equation}
where $R$ denotes the nuclear radius.
The saturation scale is  proportional to the thickness function, cf. \eqref{eq:saturationscale}, and its impact parameter dependence is implemented by
 \begin{equation}
 \begin{split}
 \bar Q^2_{s1}(\vec{x})\to  \bar Q^2_{s1}\left(\vec{x}+\frac{\vec b}{2}\right), \\
 \bar Q^2_{s2}(\vec{x})\to  \bar Q^2_{s2}\left(\vec{x}-\frac{\vec b}{2}\right).
 \end{split}
\end{equation}
The function f in \eqref{eq:CGC_simple} can then be written as
\begin{equation}
\begin{split}
    f(\vec{x}) &=  \frac{ W_0^2}{\pi Q_{s,0}^2}\frac{T_A(r_{-})T_A(r_{+})}{T_A^2(0)} \\
    &\times \left(\frac{T_A(r_{-})}{T_A(0)}\log\left(\frac{Q_{s,0}^2}{m^2}\frac{T_A(r_{+})}{T_A(0)}\right) + [r_{+}\leftrightarrow r_{-}] \right),
\end{split}
\end{equation}
where we defined $r_{\pm} = |\vec{x} \pm \vec{b}/2|$.
Remarkably, the strong coupling constant $g$ drops out. Following \cite{magma}, we will set $m=\SI{0.14}{\giga\electronvolt}$ and $Q_{s,0}=\SI{1.24}{\giga\electronvolt}$.

At this point it is worth working out some general properties of two-point models of the form
\eqref{eq:general_form} with a model-dependent function $f$.
Note that for $f(\vec{x})\propto p(\vec{x})$ this reproduces the connected part of the IPSM position space correlation function in eq.\ \eqref{eq:two_point_space_IPSM}. The function $f(\vec{x})$ changes when one transforms fields according to eq.~\eqref{eq:TwoPointFunctionChangedVariables}. 

An expression for two-mode correlation functions can be obtained from \eqref{eq:general_form} using \eqref{eq:back_transform_2point}, 
\begin{equation}
\begin{split}
 &    \left\langle \epsilon^{(m_1)}_{l_1}\epsilon^{(m_2)}_{l_2} \right\rangle =\frac{1}{2\pi c^{(m_1)}_{l_1}c^{(m_2)}_{l_2}}\int_0^\infty\mathrm{d} r \;r\; J_{m_1}\left({z^{(m_1)}_{l_1}}\rho(r)\right)\\
    &\quad\quad \times J_{m_2}\left({z^{(m_2)}_{l_2}}\rho(r)\right)\int_0^{2\pi}\frac{\mathrm{d}\phi}{2\pi}  e^{-i (m_1+m_2) \phi} f(r,\phi).
\end{split}
\end{equation}

The function $f$ can in general depend on the azimuthal angle $\phi$ in a non-trivial way, therefore it is natural to consider its Fourier expansion. Introducing $f_m(r)=1/(2\pi)\int\mathrm{d}\phi\,e^{-im\phi}f(r, \phi)$, we obtain
\begin{align}
    \left\langle \epsilon^{(m_1)}_{l_1}\epsilon^{(m_2)}_{l_2} \right\rangle
   =\sum_{m}\delta_{m_1+m_2,m} B^{(m_1,m_2;m)}_{l_1,l_2},
   \label{magma_twomode}
\end{align}
with 
\begin{equation}
\begin{split}
    B^{(m_1,m_2;m)}_{l_1,l_2} &=\frac{1}{(2\pi) c^{(m_1)}_{l_1}c^{(m_2)}_{l_2}}
    \int_0^1\mathrm{d} \rho \;\rho\; J_{m_1}\left({z^{(m_1)}_{l_1}}\rho\right)\\
    &\times J_{m_2}\left({z^{(m_2)}_{l_2}}\rho\right) \frac{f_m(\rho)}{W(\rho)}.
\end{split}
    \label{eq:magma}
\end{equation}
If the model function $f$ depends on the impact parameter vector as well, one would have to include an additional integral $\int\mathrm{d}b \, p_c(b)$ as well as an averaging integral over the reaction plane angle. In this case, however, thanks to the delta-distribution in \eqref{eq:general_form}, the reaction plane angle appears only as $\phi-\phi_R$. The $\phi_R$-average then assures that $B_{l_1,l_2}^{m_11,m_2;m}$ vanishes for $m\neq 0$.
Hence, the two-mode correlation functions vanish for $m_1+m_2\neq 0$. They are also real-valued because $B_{l_1,l_2}^{(m_1, m_2; m)}$ constitutes an integral over real-valued functions if $m=0$.

We computed two-mode correlation functions for the \textsc{Magma} model by numerically evaluating \eqref{eq:magma} with a non-symmetric background field from eq.(\ref{eq:one_point_function}) and, just like for the CGC large-$N_c$ model, by adding the contribution from geometry \eqref{eq:geometry}.

\section{Comparison of initial field models}
\label{sec:comparison}
With four initial state models at hand, we shall now compare their predicted two-mode correlations. 
We will first specify how we categorised \textsc{TrENTo} events into centrality classes.
Since the other model do not have an independent way of defining a centrality class but rather relie on the distribution of the impact parameter as an external input, we adopt the same distribution as obtained in \textsc{TrENTo} and the corresponding centrality class definition.  
Next we will compare the four different models presented before in terms of their one-point functions.
Subsequently, we will focus on two-mode correlators in the four models and compare them.

\subsection{Centrality classes}
\label{sec:CentralityClasses}
The total initial transverse entropy is to a good approximation  proportional to the final charged-particle multiplicity per unit rapidity \cite{Song2008}. This allows us to categorise \textsc{TrENTo} events in centrality classes according to multiplicity, the 0-1\% class containing those 1\% of all events with the highest multiplicity, 1-2\% referring to the succeeding 1\% and so on. The multiplicity of an event is highest in central collisions and decreases for increasing impact parameters. We will mainly constrain our analysis to two centrality classes: the rather central 0-1\% class as well as the 20-21\% class, for which we expect a non-vanishing impact parameter to play a role. Numerical data for these, as well as other centrality classes will be made available as ancillary files to this article.

Defining centrality by means of the multiplicity is useful since it is directly accessible through experiments. Another possibility is to define centrality through the impact parameter. This is the natural choice for the two CGC models, as they are parametrized by $\vec{b}$. It is not obvious, and typically not true, that the two centrality definitions lead to identical ensembles. However, in the following, we will assume ensembles to be sufficiently similar for a comparison between the different models to be valid.

\begin{figure}
    \centering
    \includegraphics[width=\columnwidth]{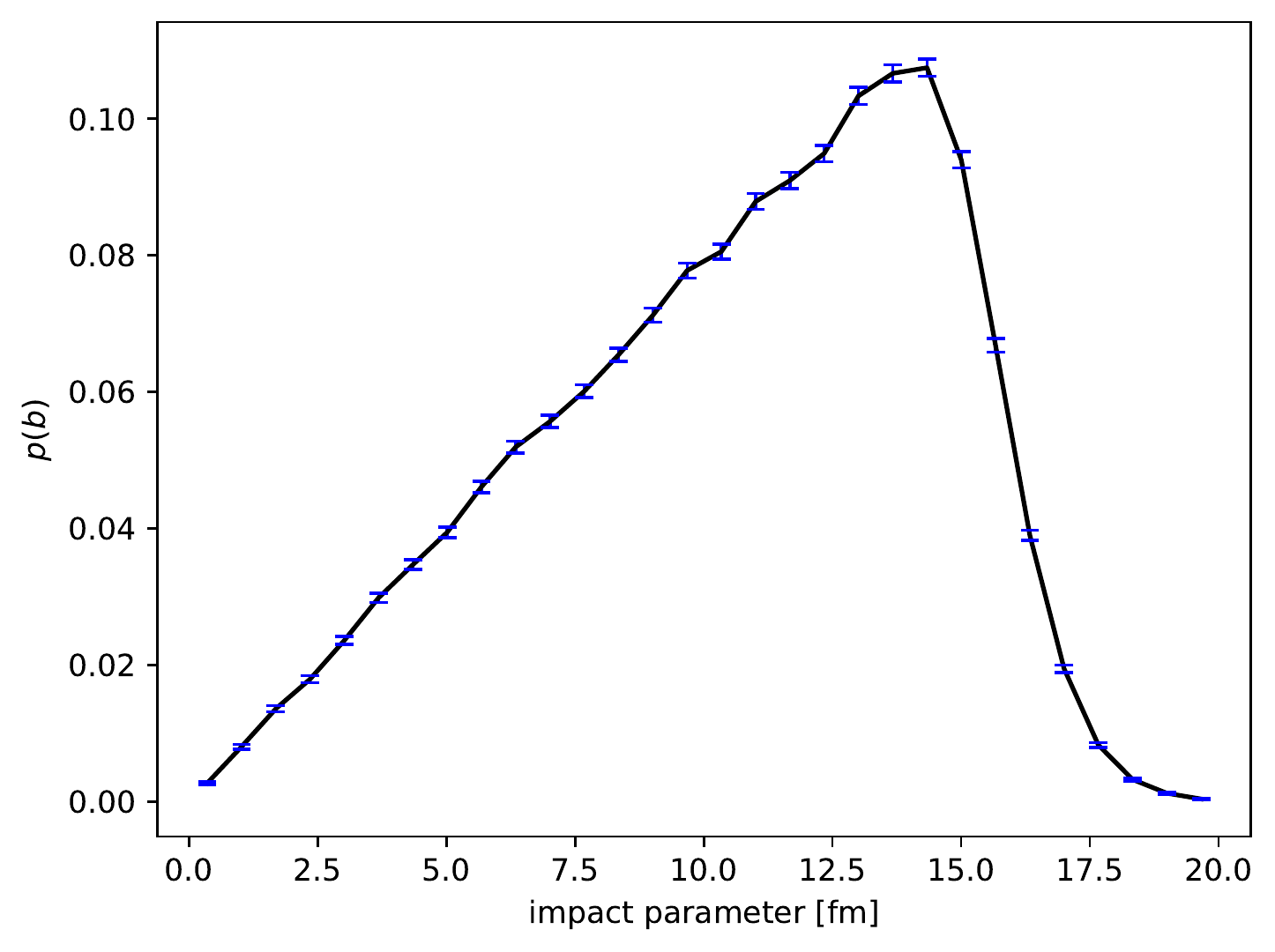}
    \caption{Impact parameter distribution with 30 histogram bins from \num{e5} Pb-Pb events generated by \textsc{TrENTo} and normalized such that $\int_0^{\infty}\mathrm{d}b\,p(b)=1$.}
    \label{fig:impact_distribution}
\end{figure}

In order to define centrality classes in terms of the impact parameter, we take the impact parameter distribution $p(b)$ obtained from \textsc{TrENTo} (cf. Fig.~\ref{fig:impact_distribution}) and use its percentiles as impact parameter window $[b_\text{min},b_\text{max}]$ of a given centrality class. The impact parameter distribution $p_c(b)$ for a centrality class of width 1\%, which enters into \eqref{eq:two_mode_LargeNc}, is then given by
\begin{equation}
    p_c(b) =
    \begin{cases}
    100\, p(b), & \text{if } b_\text{min} < b < b_\text{max}, \\
    0 & \text{otherwise}.
    \end{cases}
\end{equation}
A more sophisticated scheme would be to determine the impact parameter distribution for each (multiplicity based) centrality class with \textsc{TrENTo}.

\subsection{Expectation value of energy density}
\begin{figure*}
    \centering
    \includegraphics[width=0.37\textwidth]{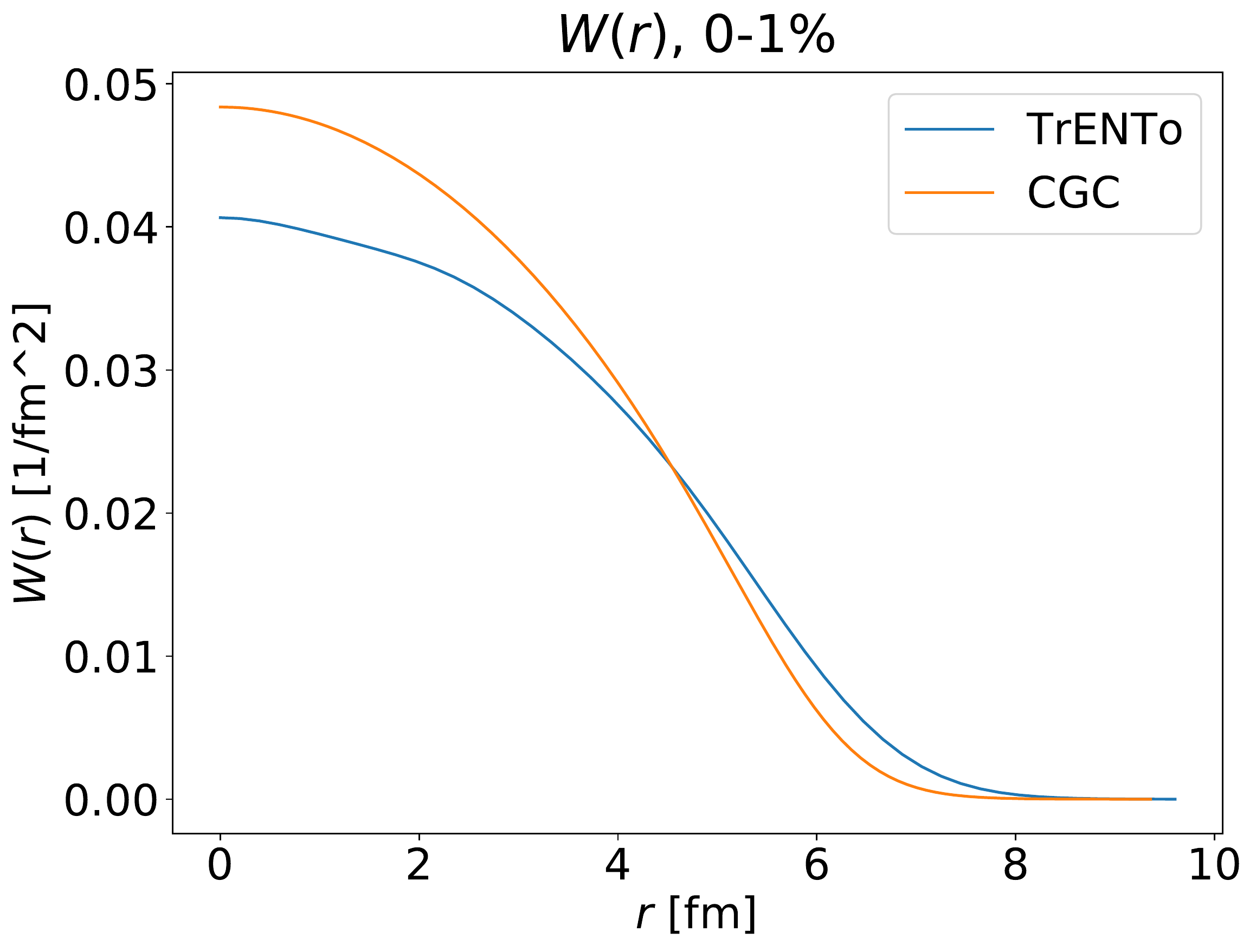} \hspace*{0.6cm}
    \includegraphics[width=0.37\textwidth]{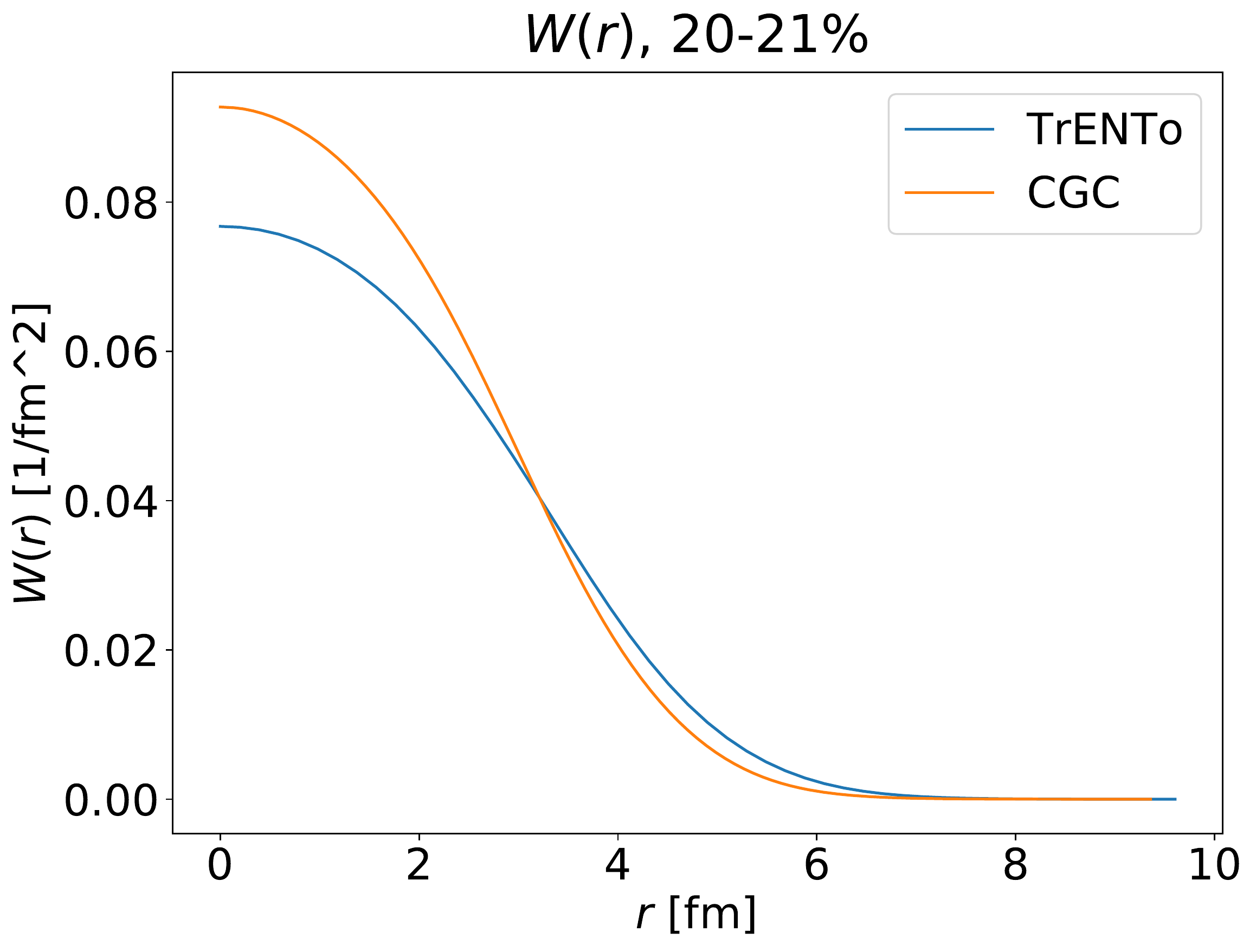}
    \caption{Background fields $W(r)$ as defined in eq.~\eqref{eq:W} for the 0-1\% centrality class (left) and the 20-21\% centrality class (right). The background field for \textsc{TrENTo} will be also used for the IPSM, while the CGC background field enters into the CGC large-$N_c$ model and \textsc{Magma}.}
    \label{fig:background_field}
\end{figure*}
The \textsc{TrENTo} model and the CGC models differ in their mean energy density definition for a given centrality class, therefore the $W$-function (defined in eq.~\eqref{eq:W}) entering in the Bessel-Fourier expansion in eq.~\eqref{eq:mode_expansion} is different for the two classes of models, as well. 
Recall that the \textsc{TrENTo} entropy density (which determines its energy density) is defined in \eqref{eq:generalized_mean} to be proportional to a generalized $p$-average of the reduced thickness functions. In contrast, the energy density in the CGC models  is given by \eqref{eq:CGCenergy} as the product of the saturation scales $ Q^2$ of two nuclei, the position dependence of which is taken proportional to the reduced thickness functions of the corresponding nuclei.
The \textsc{Magma} model and the CGC large-$N_c$ share the same energy density definition. 
In the IPSM model the energy density can be considered as an input, therefore we choose to take \textsc{TrENTo} one-point functions.  

In Fig. \ref{fig:background_field} we show the resulting background field profiles $W(r)$ as function of the radius in the two classes of models, for the 0-1\% and the 20-21\% classes.  The \textsc{TrENTo} profile is slightly broader than the CGC energy density on average, which is due to different definitions of energy density in terms of thickness function eq. \eqref{eq:generalized_mean} with $p=0$ in \textsc{TrENTo} and eq. \eqref{eq:CGCenergy} for the CGC models. 

\subsection{One-point functions}
The IPSM necessitates the coefficients $\bar{\epsilon}_l^{(m)}$ in order to predict two-mode correlations (cf.~eq.~\eqref{eq:two_point_IPSM_random_general}). They also appear in the geometry term \eqref{eq:geometry}, which has to be added to the CGC large-$N_c$ and \textsc{Magma} model to compute two-mode functions with randomized reaction angles. 
As has been discussed in section \ref{sec:Geometry}, the coefficients $\bar{\epsilon}_l^{(m)}$ are related to expectation values $\langle \epsilon_l^{(m)}\rangle$ at a fixed reaction plane angle $\phi_R=0$. 

\begin{figure*}
\centering
\includegraphics[width=0.37\textwidth]{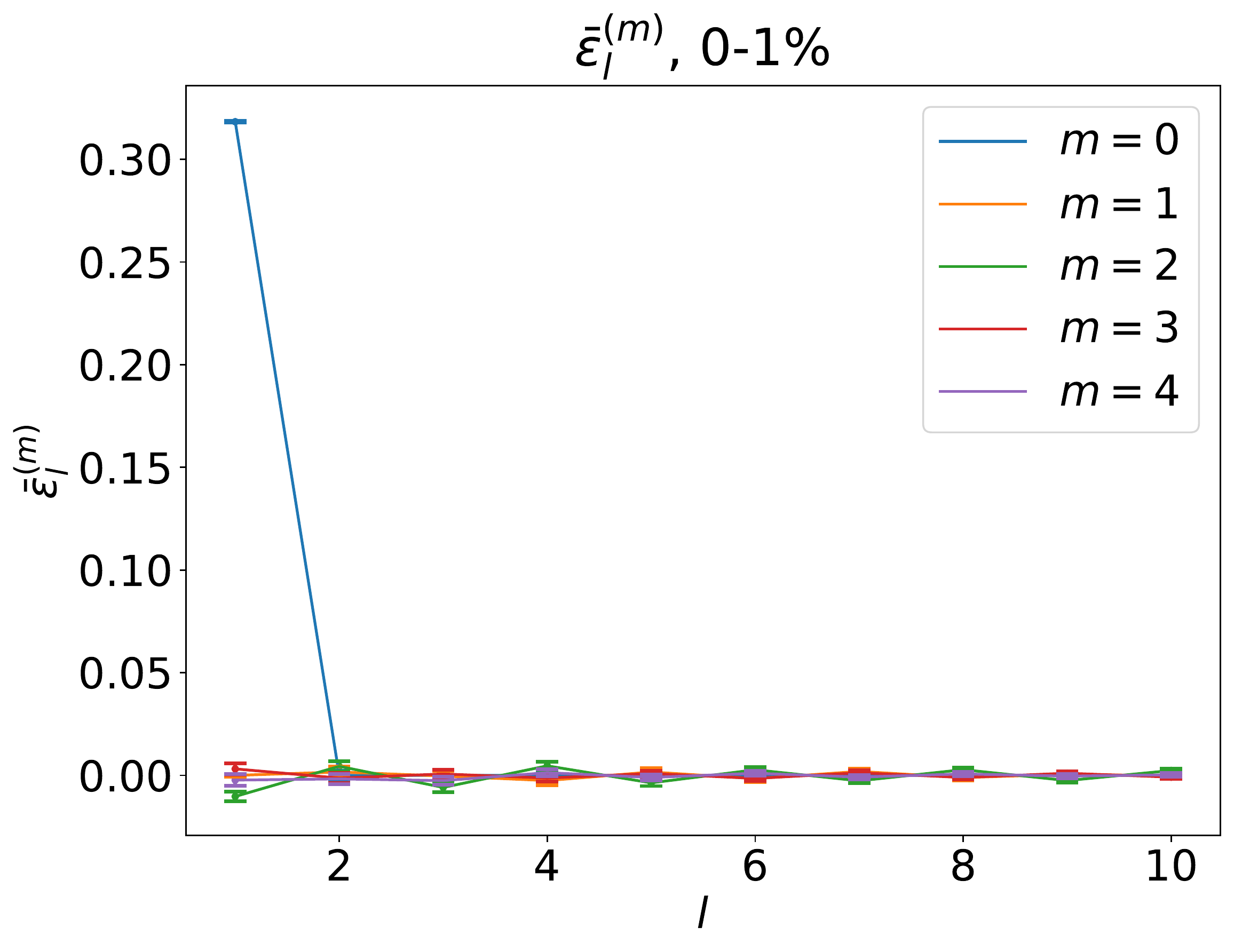} \hspace*{0.6cm}
\includegraphics[width=0.373\textwidth]{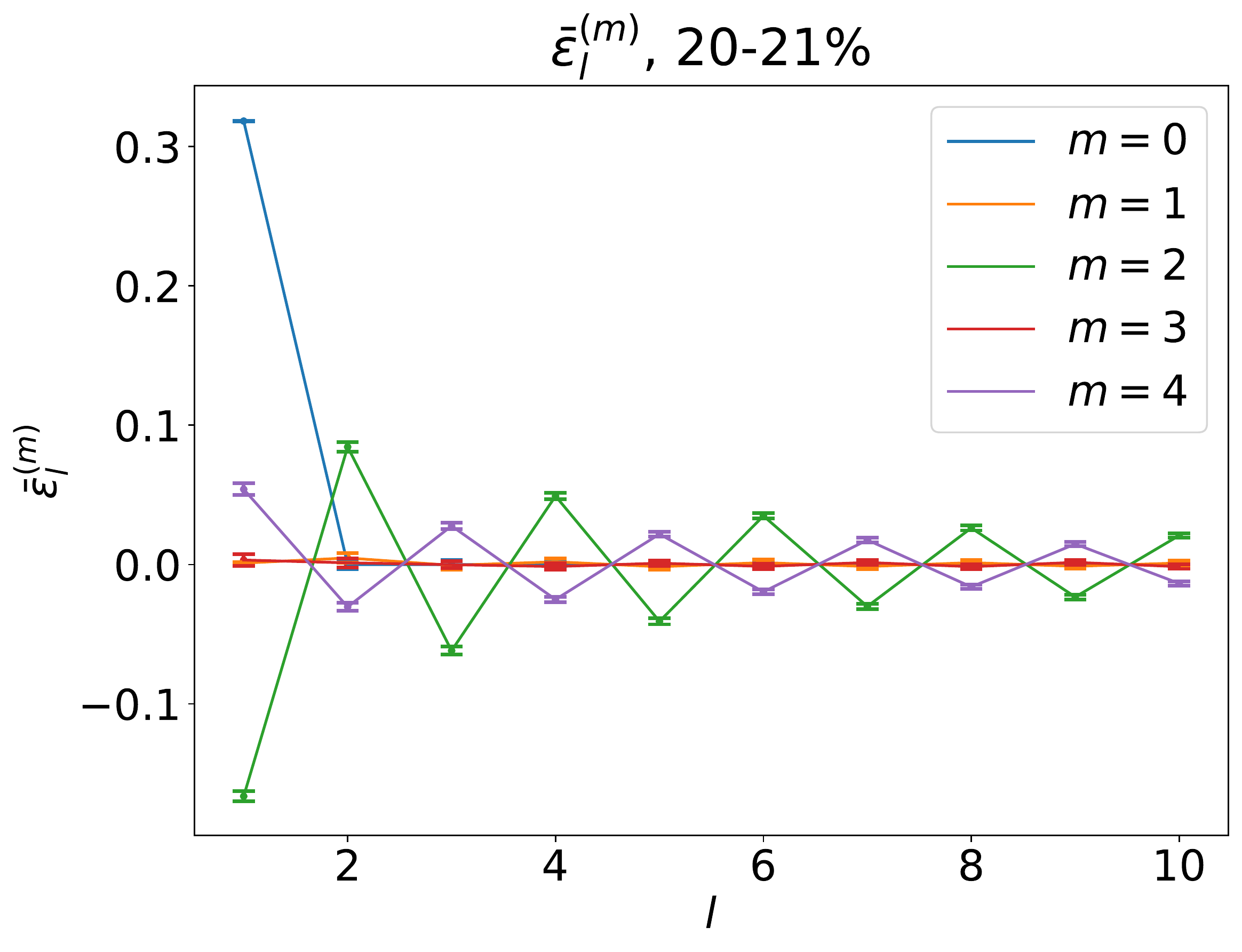}
\caption{Coefficients $\bar{\epsilon}_l^{(m)}$ as defined in eq.\ \eqref{eq:expValueEpsilonNonSym} for the 0-1\% centrality class (left) and 20-21\% centrality class (right), obtained from \num{e5} Pb-Pb events generated by \textsc{TrENTo}. The error bars correspond to the statistical uncertainty of the respective means. Lines have been added to guide the eye.}
\label{fig:one_point_trento}
\end{figure*}

In Fig.~\ref{fig:one_point_trento} we present $\bar{\epsilon}_l^{(m)}$ computed from \textsc{TrENTo} events as a function of $l$ and $m$ for two centrality classes. For both centrality classes, expectation values with $m$ odd vanish, as follows from symmetry considerations. In addition, for $m=0$, only $\bar{\epsilon}_1^{(0)}=1/\pi\approx 0.32$ is non-vanishing, as we have concluded before. 

Note also that the moduli of the non-vanishing expectation values  decay approximately exponentially with increasing values of $l$, which highlights \emph{a posteriori} that our chosen set of basis functions is well suited to describe initial field profiles using a low number of expansion coefficients. We note that the decay as a function of $l$ is much more quick in the 0-1\% class than in the 20-21\% class. This is only natural as the coefficients $\bar{\epsilon}_l^{(m)}$ quantify azimuthal variations of the background field, cf. \eqref{eq:mode_expansion_p}. The background field, on the other hand, is symmetric for central collisions and ever more elliptically deformed for higher centrality classes.

\begin{figure*}
\centering
\includegraphics[width=0.7\textwidth]{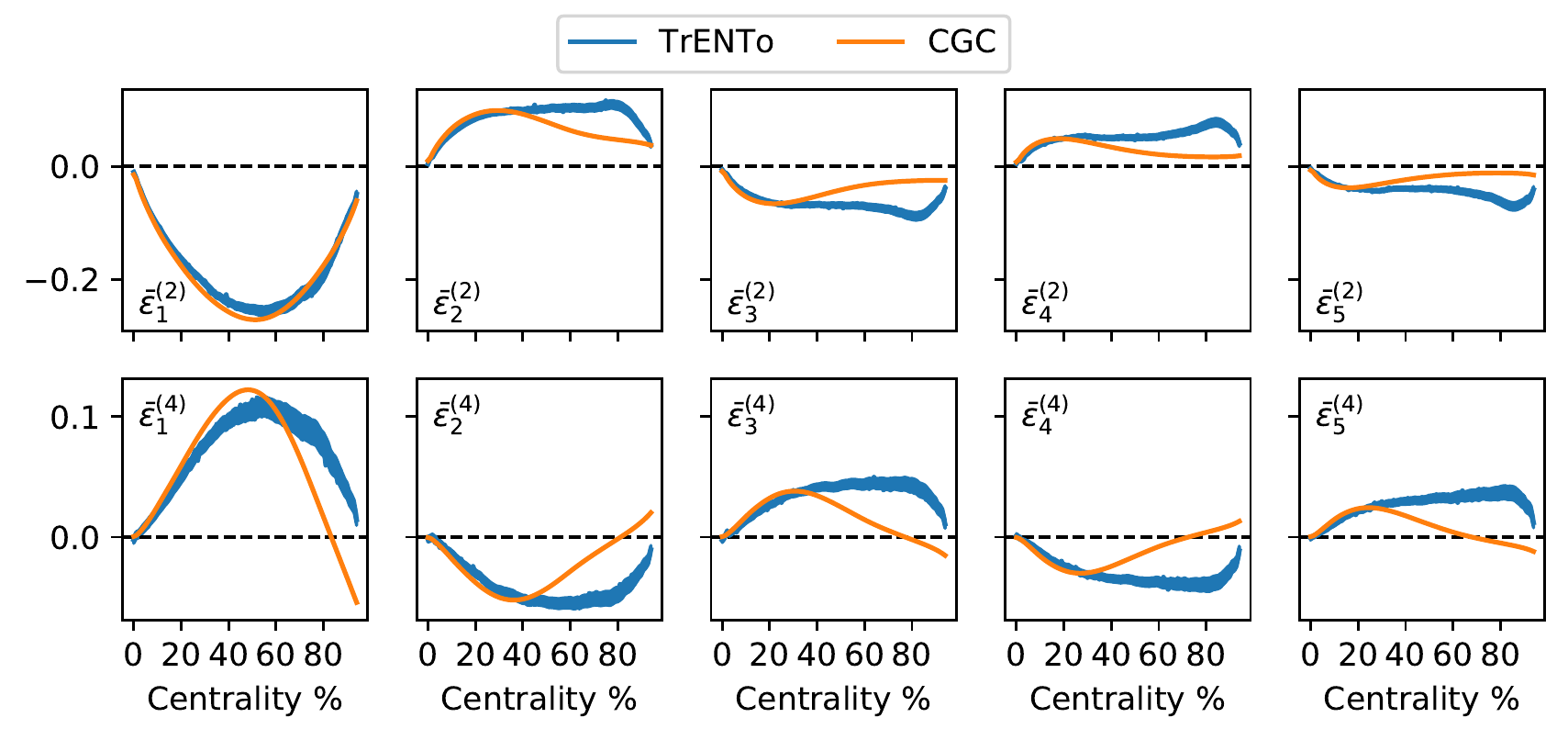}
\caption{Comparison of the coefficients $\bar{\epsilon}_l^{(m)}$ as defined in eq.\ \eqref{eq:expValueEpsilonNonSym} from the \textsc{TrENTo} model and CGC, as a function of centrality. \textsc{TrENTo} data obtained from \num{e5} Pb-Pb events. Line thickness for \textsc{TrENTo} reflects the statistical uncertainty.}
\label{fig:one_point_comparison}
\end{figure*}

In Fig.~\ref{fig:one_point_comparison} we compare the coefficients $\bar{\epsilon}_l^{(m)}$ between \textsc{TrENTo} and CGC as function of centrality. 
The two models agree with each other for all $m$ and $l$ for central collisions. On the other hand, for peripheral collisions energy density fluctuations differ substantially between the two models.
This can be seen as a consequence of the different definitions of the energy density in terms of the reduced thickness functions. 

The coefficients $\bar{\epsilon}_l^{(m)}$ as defined in \eqref{eq:expValueEpsilonNonSym} and displayed in Fig.\ \ref{fig:one_point_trento} can fully characterize the contribution to harmonic flow coefficients from the collision geometry. To characterize contributions from fluctuations we need also two-point functions, to which we turn next.

\subsection{Two-mode correlation functions}
We shall now turn to our key quantity of interest, namely connected two-mode correlation functions of ensembles with randomized reaction plane angles. 
For this discussion we introduce the following notation
\begin{equation}
G_{l_1, l_2}^{(m_1, m_2)} = \left \langle \epsilon_{l_1}^{(m_1)} \epsilon_{l_2}^{(m_2)} \right \rangle_{\circ, c},
\end{equation}
for the connected two-mode correlation function, and for the diagonal part
\begin{equation}
G_{l}^{(m)} = G_{l, l}^{(m, -m)}.
\label{eq:defDiagonalCorrelator}
\end{equation}
It follows from the statistical azimuthal rotation symmetry that 
the two-mode correlation functions are real-valued and vanish except for $m_1+m_2=0$. 
We can therefore constrain our analysis to real-valued correlators of the form $G_{l_1,l_2}^{(m, -m)}$. 

We have explained in the previous section for each individual model how to retrieve two-mode correlators. However, we still need to fix the constants $\alpha$ and $\beta$ in the IPSM for comparisons to the other models to be possible. This was carried out by fitting the IPSM expression \eqref{eq:two_point_IPSM_random_general} to five \textsc{TrENTo} correlators, namely those $G_1^{(m)}$ with $0 \leq m \leq 4$. There are two reasons for this specific choice. Firstly, we demanded correlators on the diagonal as we can expect them to be non-vanishing regardless of centrality. And secondly, we made sure to choose correlators with a small index $l$. The reason is that we cannot expect the IPSM with its point-shaped contributions to remain valid as we pass to finer details in position space. The best-fit values for $\alpha$ and $\beta$ for the two centrality classes in question are presented in Tab. \ref{tab:IPSM_params}.
\begin{table}
    \centering
    \caption{Best-fit parameter values of the IPSM. Values obtained by a least-squares minimization to the five \textsc{TrENTo} data points $G_1^{(m)}, 0 \leq m \leq 4$.}
    \begin{ruledtabular}
    \begin{tabular}{ccc}
    class & $\alpha$ & $\beta$ \\
    \hline
    0-1\% &  \num{6.1e-3} &  \num{4.3e-3} \\
    20-21\% & \num{1.3e-2} & \num{3.2e-3} \\
    \end{tabular}
    \end{ruledtabular}
    \label{tab:IPSM_params}
\end{table}

\begin{figure*}
\centering
\includegraphics[width=0.8\textwidth]{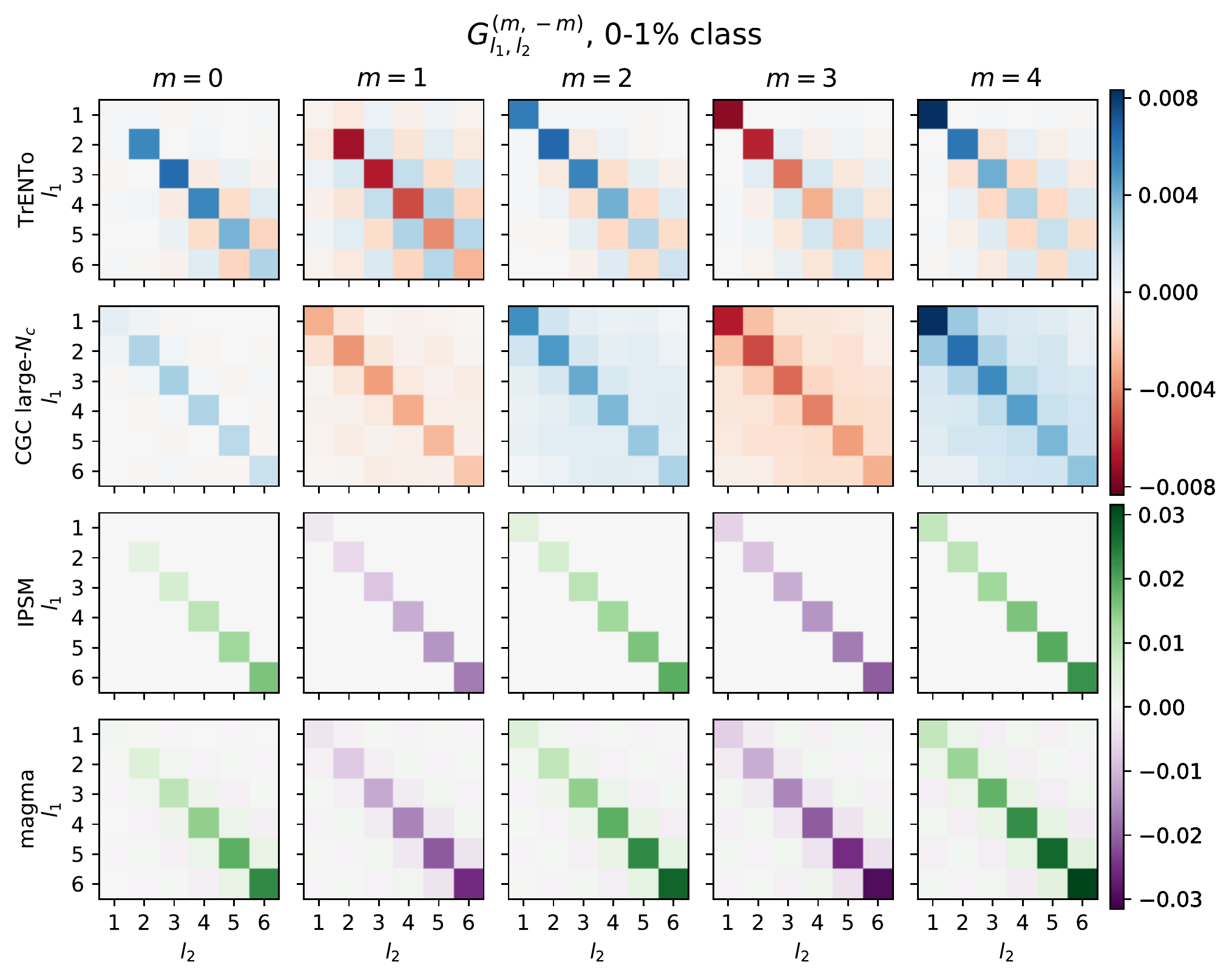}
\caption{Connected two-mode correlation functions with $m_1+m_2=0$ for four initial-state models, averaged over the 0-1\% centrality class. Different colors are used to distinguish positive and negative values as well as the two colorbar scales.}
\label{eq:two_comparison_central}
\end{figure*}
\begin{figure*}
\centering
\includegraphics[width=0.8\textwidth]{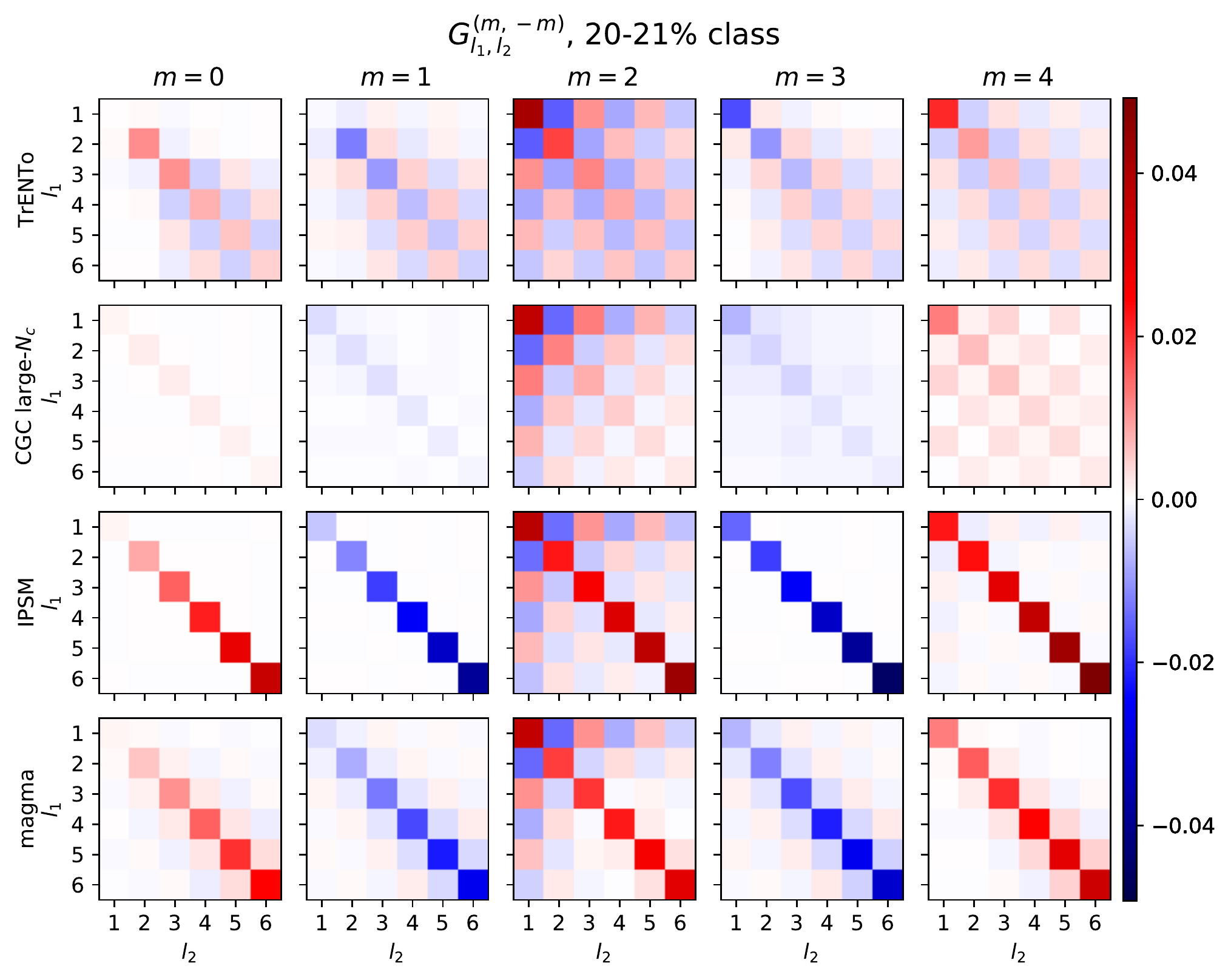}
\caption{Connected two-mode correlation functions with $m_1+m_2=0$ for four initial-state models, averaged over the 20-21\% centrality class. Different colors are used to distinguish positive and negative values.}
\label{eq:two_comparison_peripheral}
\end{figure*}

In Figs.~\ref{eq:two_comparison_central} and \ref{eq:two_comparison_peripheral} we present color plots of $G_{l_1 l_2}^{(m, -m)}$, with azimuthal wave numbers $m = 0, 1, 2, 3, 4$, and as a function of the radial wave numbers $l_1$ and $l_2$ for the two centrality classes 0-1\% and 20-21\% and comparing the four models introduced before. Note that this representation is only possible because the correlators are real-valued and depend for fixed $m$ only on the two indices $l_1, l_2$.

In the 0-1\% class, all four models show approximately diagonal two-mode correlation functions. While the off-diagonal correlators are almost rigorously zero in the IPSM, they are more pronounced in \textsc{TrENTo}, especially for higher values of $l_1$, $l_2$.

Both the IPSM and the \textsc{Magma} model have significantly larger diagonal values $G_l^{(m)}$ than the \textsc{TrENTo} and CGC large-$N_c$ model. This is actually the reason to use two different color schemes in Fig.~\ref{eq:two_comparison_central}. Moreover, one observes here already that the diagonal values $G_l^{(m)}$ decay with $l$ in the \textsc{TrENTo} and CGC large-$N_c$ model, while they actually increase for the IPSM and \textsc{Magma} models. This is directly linked to the assumption of point-like sources. We will further comment on this below.

The off-diagonal elements in the two CGC models are -- relative to the corresponding diagonal elements -- smaller than they are in \textsc{TrENTo}. In a direct comparison of the two CGC models, the correlators of the \textsc{Magma} model are in this sense more diagonal than those of the CGC large-$N_c$ model.

In the IPSM, strict diagonality is a direct consequence of centrality: The coefficients $\bar{\epsilon}_l^{(m)}$ vanish for $(m, l) \neq (0,1)$ in central events so that only the term proportional to $\delta_{l_1,l_2}$ in expression \eqref{eq:two_point_IPSM_random_general} survives. In constrast to the IPSM, the off-diagonal elements in \textsc{TrENTo} are likely to result from the fact that the assumption of \emph{point-shaped} sources is dropped in favour of an extended Gaussian shape. This fits with the observation that the off-diagonal elements become more pronounced for higher radial wavenumbers $l, l'$, which probe finer structures in position space. 

When passing from the 0-1\% class (Fig.\ \ref{eq:two_comparison_central}) to the 20-21\% class (Fig.\ \ref{eq:two_comparison_peripheral}), one can equally make out deviations from the initially diagonal structure, most notably for those entries with either $l_1=1$ or $l_2=1$ of color plots with $m=2$ and (less prominently) $m=4$. These off-diagonal contributions, however, result from non-centrality of the collisions, expressed through the geometry term in eq.\ \eqref{eq:geometry}.

A further aspect that all color plots share, concerns the sign of the diagonal elements. Indeed we have
\begin{equation}
   \text{sign}\left(G_{l}^{(m)}\right) = (-1)^{m}.
   \label{eq:sign_diagonal}
\end{equation}
This a direct consequence of eq.~\eqref{eq:neg_coeff} which implies
\begin{equation}
\begin{split}
    G_l^{(m)} &= \left\langle \epsilon_l^{(m)}\epsilon_l^{(-m)} \right\rangle = (-1)^{m} \left\langle \epsilon_l^{(m)}\epsilon_l^{(m)*} \right\rangle \\ 
    &= (-1)^{m} \left\langle |\epsilon_l^{(m)}|^2 \right\rangle.
    \end{split}
\end{equation}

The plots establish that two-mode correlation functions are largest on the diagonal. On the other hand, the correlator $G_1^{(0)}$ is vanishing in all models. It corresponds to the variance of $\epsilon_1^{(0)}$, which quantifies fluctuations of the integrated field, i.e. total transverse energy. Fluctuations of this quantity are naturally suppressed for \emph{narrow} centrality classes, as we consider them here.

\begin{figure*}
\centering
\includegraphics[width=0.7\textwidth]{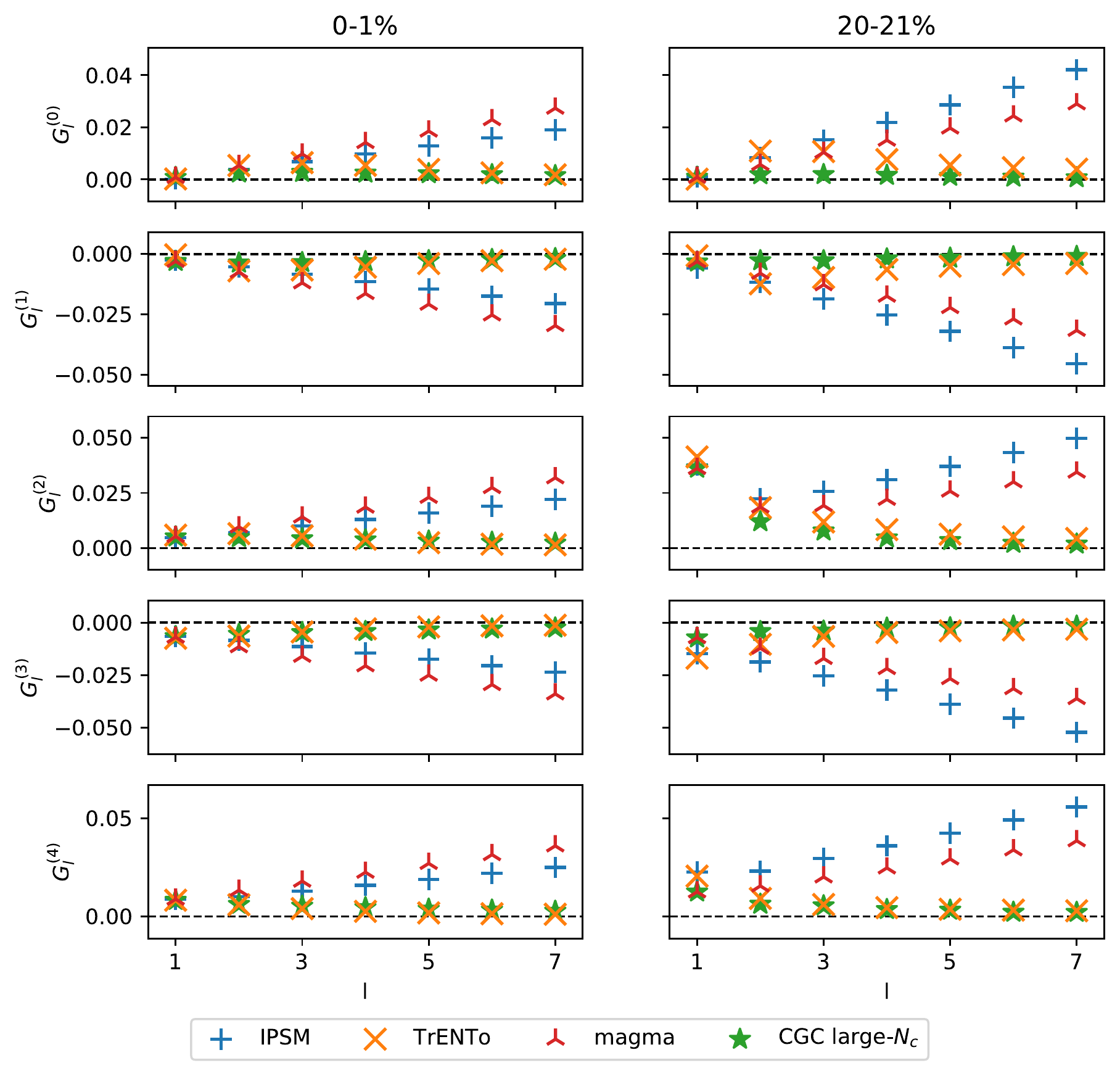}
\caption{Variation of the diagonal values of two-point correlators $G_{l}^{(m)}$ (defined in eq.\ \eqref{eq:defDiagonalCorrelator}) as a function of the radial wave number $l$ in the four models for two centrality classes in comparison. The left and right column correspond to the 0-1\% class and the 20-21\% class, respectively. See text for further discussion.}
\label{fig:IPSM_Trento}
\end{figure*}

In Fig.~\ref{fig:IPSM_Trento} we compare the variation of the correlators $G_l^{(m)}$ on the diagonal  as a function of $l$ in the four models and for different values of $m$. As we established before, the sign of the data series follows a $(-1)^m$-pattern. Regardless of centrality, the diagonal elements in \textsc{TrENTo} and the CGC large-$N_c$ model seem to converge towards zero for large radial wave numbers $l$ after peaking at around $l\approx 3$. For $m=2$, the geometry part covers the peak in the 20-21\% class, making the curve decrease monotonously. In contrast, the correlators of the IPSM and \textsc{Magma} diverge linearly with $l$ and with similar slopes. This is likely a consequence of the point-like approximation for correlations underlying these two models. It is well possible that the higher $l$ modes are actually efficiently damped by a viscous fluid evolution and that the increasing behaviour for large $l$ is therefore not visible in final state observables. This will be investigated in further work.

Interestingly, the \textsc{TrENTo} correlators agree fairly well with those of the CGC large-$N_c$ model, in both centrality classes. For small values of $l$, the IPSM agrees with the previous two models as well, only from around $l=3$ on can an increasing disparity be observed. Similarly, \textsc{Magma} data matches the other models up to $l=3$ for $m=0$ and $l=1$ for $m=4$, while diverging away for higher values of $l$.

The variation of the IPSM correlators shows that the assumption of point-shaped sources is fairly valid as long as one probes coarse structures in position space. However, for radial modes with large wave numbers $l$, the finer structure of correlations in position space is unveiled. Hence, the IPSM begins to differ from models working with finite source extensions. The close similarity between the IPSM and \textsc{Magma} is natural, because the latter model can be seen a  generalization of the IPSM but still shares with it a Dirac-distribution-shaped contact term.

\begin{figure*}
\centering
\includegraphics[width=0.76\textwidth]{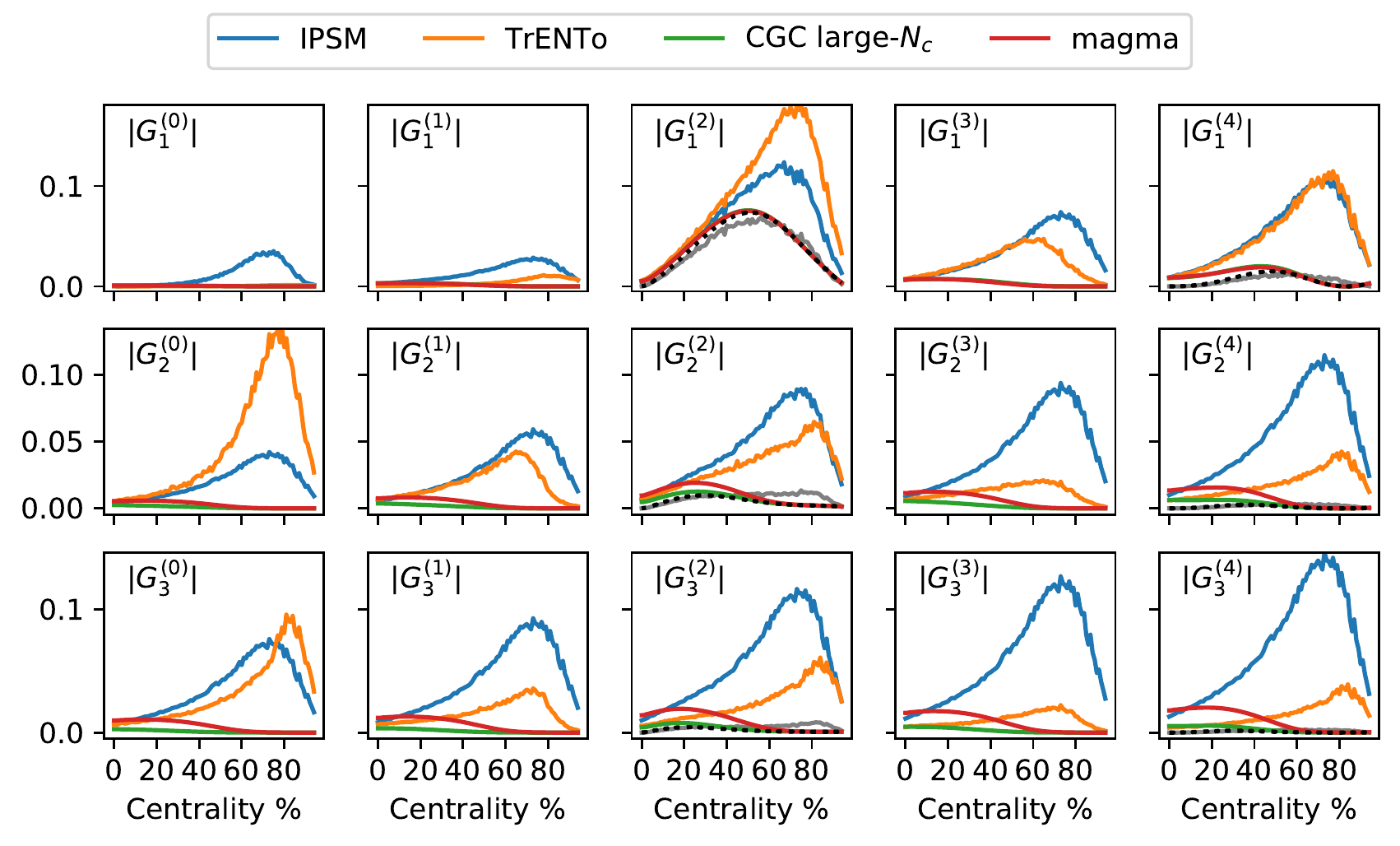}
\caption{Variation of $|G_{l}^{(m)}|$ in the four models as a function of centrality. The dotted black lines for $m=2$ and $m=4$ correspond to the contribution from geometry which has been added to the two CGC models. In comparison, the solid grey lines stand for the corresponding geometry contribution in \textsc{TrENTo}. The sign of a correlator $G_l^{(m)}$ can be recovered from eq.~\eqref{eq:sign_diagonal}.}
\label{fig:correlators_centrality}
\end{figure*}

Finally in Fig.\ \ref{fig:correlators_centrality} we show how the diagonal parts of the two-mode correlators $G_l^{(m)}$ (defined in eq.\ \eqref{eq:defDiagonalCorrelator}) depend on centrality. 
The absolute value of the presented correlators seems to be monotonously increasing as a function of centrality. This is a natural consequence of geometry contributing more significantly with rising centrality to the correlators.

As for the agreement of the four models with each other, this can be investigated with respect to the mode numbers $m$ and $l$ as well as centrality.
Fig.~\ref{fig:correlators_centrality} suggests that the four models lead to similar results for central collisions and show increasing discrepancy as a function of centrality. 

Generally, for low values of $l$ and $m$, the four models lead to similar results, the discrepancy depending only slightly on centrality. Instead, the models diverge faster away from each as a function of centrality when increasing values of $m$ and $l$ are chosen. The $m$-dependence of this phenomenon seems weaker than its $l$-dependence. In addition, the \textsc{TrENTo} data matches up well with the large-$N_c$ model, and the IPSM with \textsc{Magma}, as we discussed before. Equally, the $l$-dependence supports our previous discussion of the limits of the IPSM resulting from the point-shaped nature of its sources. 

We note also that \textsc{TrENTo} and the IPSM show in all panels of Fig.~\ref{fig:correlators_centrality} a maximum  around the $80 \% $ centrality. It is difficult to give meaning to the precise position of this maximum, but it is also related to the specific choice of the function $W(r)$ in the definition of the mode expansion in eq.\ \eqref{eq:mode_expansion}. While the latter is not unique, the actual correlation functions of energy density in position space are independent of this choice.

Interestingly the geometry contribution of the CGC large-$N_c$ model dominates the correlation function (dotted line respect to green line in Fig.~\ref{fig:correlators_centrality}), meanwhile the corresponding contribution in \textsc{TrENTo} (grey line)  is much smaller than the value of the two point function (orange line). 

Fig.~\ref{fig:correlators_centrality} thus suggests that the four models agree fairly well with each other for sufficiently low radial wave numbers $l$ and centrality classes. In this regime, the analytically solvable independent point source model (IPSM) would then exhibit \emph{universal} properties shared with other initial state models. A detailed investigation of our results shows that this universality is restricted to not too large values of the radial wave number $l$ and more pronounced for the more central multiplicity classes. This is interesting because one expects that higher $l$ modes are anyway damped more strongly by dissipative effects in the fluid regime. 

%
%

\section{Conclusion and outlook}
\label{sec:conclusions}
We have studied here models for the event-by-event fluctuations in the initial energy density distributions of Pb-Pb collisions at $\sqrt{s_{\text{NN}}}=\SI{2.76}{\tera\electronvolt}$ in ensembles with randomized reaction plane angles. For this purpose, we used a complete Fourier-Bessel mode expansion scheme for fluctuations around a background profile specific to a given centrality class. 
The statistical properties of different models for the initial state can then be characterized through one-mode and two-mode correlation functions depending on azimuthal wave numbers $m$ and radial wave numbers $l$.

Comparing four different initial state models that are currently discussed in the 
literature, we have computed two-mode correlation functions of initial energy density. While \textsc{TrENTo} and the two Color Glass Condensate (CGC) models demanded a numerical evaluation, the independent point-sources model (IPSM) allowed to find analytical expressions for correlators. Presenting the results exemplary for the centrality classes 0-1\% and 20-21\%, we were able to capture characteristics of fluctuations for central collisions as well as effects of the collision geometry for less central collisions.

Indeed, the IPSM qualitatively agrees with three significantly more extended initial state models, as non-vanishing two-mode correlation functions are of approximately diagonal form with off-diagonal corrections in non-central collisions. For small radial wave numbers $l$, the IPSM agrees even quantitatively with the other models, while differences could be explained with the point-shaped nature of sources in the IPSM for larger $l$.

An important outcome of this study are actually the concrete values of one-point functions and two-mode correlators for the different models. We have obtained them in numerical form and will make them available to the public as ancillary files accompanying this article. In a future publication we plan to use these initial data together with the \textsc{FluiduM} framework \cite{Floerchinger:2013vua,Floerchinger:2013hza,Floerchinger2014,Floerchinger2014mode,Floerchinger:2017cii,Mazeliauskas:2018irt,Floerchinger:2018pje,Devetak:2019lsk}. This will allow to calculate from them two-particle correlation functions and harmonic flow coefficients that can be measured experimentally. We are curious to see whether any of the four models discussed here is favoured by experimental data. For a first step one could actually use the fluid parameters (overall normalization of entropy density, initialisation time, viscosities and freeze-out temperature) obtained in ref. \cite{Devetak:2019lsk} by fitting to transverse momentum spectra of identified particles. No additional parameters would be needed and one could directly see which initial state model works best.

An interesting possibility also arises from the observation that the two-mode correlation functions of all four investigated models agree reasonably well for small radial wave numbers $l$ and for central collisions, indicating there a form of universality. It is expected that modes with higher values of $l$ are damped more strongly by shear and bulk viscous dissipation. Also they lead to oscillating patterns on the freeze-out surface and one can therefore expect that they have a weaker contribution to final state two-particle correlation functions. The observed universality for central collisions suggests to concentrate on those in order to constrain thermodynamic and transport properties of the quark-gluon plasma, while more peripheral centrality classes might be more suitable to distinguish between different initial state models (see also \cite{Giacalone:2017uqx,Teaney:2010vd,Teaney:2013dta}).

Before closing, we would like to mention possible ways to continue this line of research. 
One could  extend the analysis in this paper to $n$-mode correlation functions with $n>2$.  Furthermore, one should also look into other collision systems than Pb-Pb. 
The general methods employed here can be applied to arbitrary collision systems as long as a fluid description is applicable.

\begin{acknowledgments}
The authors thank G.~Giacalone for helpful discussions and comments. They also thank S.~Masciocchi, A.~Mazeliauskas and other members of the ISOQUANT collaboration for useful discussions.
This work is part of and supported by the Deutsche Forschungsgemeinschaft (DFG, German Research Foundation) Collaborative Research Centre “SFB 1225 (ISOQUANT)” as well as project FL 736/3-1.  E.~G.\ is supported by the U.S. Department of Energy, Office of Science, Office of Nuclear Physics, grants Nos. DE\nobreakdash-FG\nobreakdash-02\nobreakdash-08ER41450.
\end{acknowledgments}

\bibliography{bibliography}

\end{document}